\documentclass[12pt]{article}
\usepackage{amssymb}

\def\hybrid{\topmargin 0pt      \oddsidemargin 0pt
        \headheight 0pt \headsep 0pt
        \voffset=-0.5cm
        \hoffset=-0.25in
        \textwidth 6.75in
        \textheight 9.5in       
        \marginparwidth 0.0in
        \parskip 5pt plus 1pt   \jot = 1.5ex}
\catcode`\@=11
\def\marginnote#1{}

\newcount\hour
\newcount\minute
\newtoks\amorpm
\hour=\time\divide\hour by60 \minute=\time{\multiply\hour by60 \global\advance\minute by-\hour}
\edef\standardtime{{\ifnum\hour<12 \global\amorpm={am}%
        \else\global\amorpm={pm}\advance\hour by-12 \fi
        \ifnum\hour=0 \hour=12 \fi
        \number\hour:\ifnum\minute<10 0\fi\number\minute\the\amorpm}}
\edef\militarytime{\number\hour:\ifnum\minute<10 0\fi\number\minute}

\def\draftlabel#1{{\@bsphack\if@filesw {\let\thepage\relax
   \xdef\@gtempa{\write\@auxout{\string
      \newlabel{#1}{{\@currentlabel}{\thepage}}}}}\@gtempa
   \if@nobreak \ifvmode\nobreak\fi\fi\fi\@esphack}
        \gdef\@eqnlabel{#1}}
\def\@eqnlabel{}
\def\@vacuum{}
\def\draftmarginnote#1{\marginpar{\raggedright\scriptsize\tt#1}}
\def\draftlabel#1{{\@bsphack\if@filesw {\let\thepage\relax
   \xdef\@gtempa{\write\@auxout{\string
      \newlabel{#1}{{\@currentlabel}{\thepage}}}}}\@gtempa
   \if@nobreak \ifvmode\nobreak\fi\fi\fi\@esphack}
        \gdef\@eqnlabel{#1}}
\def\@eqnlabel{}
\def\@vacuum{}
\def\draftmarginnote#1{\marginpar{\raggedright\scriptsize\tt#1}}

\def\draft{\oddsidemargin -.5truein
        \def\@oddfoot{\sl preliminary draft \hfil
        \rm\thepage\hfil\sl\today\quad\militarytime}
        \let\@evenfoot\@oddfoot \overfullrule 3pt
        \let\label=\draftlabel
        \let\marginnote=\draftmarginnote
   \def\@eqnnum{(\theequation)\rlap{\kern\marginparsep\tt\@eqnlabel}%
\global\let\@eqnlabel\@vacuum}  }

\def\numberbysection{\@addtoreset{equation}{section}
        \def\theequation{\thesection.\arabic{equation}}}

\def\underline#1{\relax\ifmmode\@@underline#1\else
        $\@@underline{\hbox{#1}}$\relax\fi}

\def\titlepage{\@restonecolfalse\if@twocolumn\@restonecoltrue\onecolumn
     \else \newpage \fi \thispagestyle{empty}\c@page\z@
        \def\thefootnote{\fnsymbol{footnote}} }

\def\endtitlepage{\if@restonecol\twocolumn \else  \fi
        \def\thefootnote{\arabic{footnote}}
        \setcounter{footnote}{0}}  

\numberbysection \hybrid

\newcounter{mo}

\newcommand{\tr}{{\rm tr}}
\newcommand{\ti}[1]{\tilde{#1}}

\newcommand{\vf}{\varphi}
\newcommand{\al}{\alpha}
\newcommand{\be}{\beta}
\newcommand{\ga}{\gamma}
\newcommand{\om}{\omega}
\newcommand{\vth}{\vartheta}

\newcommand{\Mat}{ {\rm Mat}(N,\mathbb C) }
\newcommand{\Matm}{ {\rm Mat}(M,\mathbb C) }
\newcommand{\mR}{\mathfrak R}

\newcommand{\hS}{ {\hat S} }
\newcommand{\hL}{ {\hat L} }

\newtheorem{predl}{Proposition}[section]

\def\beq{\begin{equation}}
\def\eq{\end{equation}}
\def\p{\partial}

\def\res{\mathop{\hbox{Res}}\limits}

\begin{document}

\setcounter{page}{1}

\date{}
\date{}
\vspace{50mm}

\begin{flushright}
 ITEP-TH-06/15\\
\end{flushright}
\vspace{0mm}

\begin{center}
\vspace{8mm}
%
{\LARGE{Yang-Baxter equations with two Planck constants}}
\\
\vspace{12mm} {\large {A. Levin}\,$^{\flat\,\sharp}$ \ \ \ \ {M.
Olshanetsky}\,$^{\sharp\,\ddagger\,\natural}$
 \ \ \ {A. Zotov}\,$^{\diamondsuit\, \sharp\, \natural}$ }\\
 \vspace{8mm}

 \vspace{2mm} $^\flat$ -- {\small{\sf 
 NRU HSE, Department of Mathematics,
 Myasnitskaya str. 20,  Moscow,  101000,  Russia}}\\
 \vspace{2mm} $^\sharp$ -- {\small{\sf 
 ITEP, B. Cheremushkinskaya str. 25,  Moscow, 117218, Russia}}\\
 \vspace{2mm} $^\natural$ -- {\small{\sf MIPT, Inststitutskii per.  9, Dolgoprudny,
 Moscow region, 141700, Russia}}\\
\vspace{2mm} $^\ddagger$ -- {\small{\sf
IITP (Kharkevich Institute) RAS,
Bolshoy Karetny per. 19, Moscow, 127994,  Russia}}\\
\vspace{2mm} $^\diamondsuit$ -- {\small{\sf Steklov Mathematical
Institute  RAS, Gubkina str. 8, Moscow, 119991,  Russia}}\\
\end{center}

\begin{center}\footnotesize{{\rm E-mails:}{\rm\ \
 alevin@hse.ru,\  olshanet@itep.ru,\  zotov@mi.ras.ru}}\end{center}

 \begin{abstract}
We consider Yang-Baxter equations arising from its associative
analog and study corresponding exchange relations. They generate
finite-dimensional quantum algebras which have form of coupled ${\rm
GL}(N)$ Sklyanin elliptic algebras. Then we proceed to a natural
generalization of the Baxter-Belavin quantum $R$-matrix to the case
${\rm Mat}(N,\mathbb C)^{\otimes 2}\otimes {\rm Mat}(M,\mathbb
C)^{\otimes 2}$. It can be viewed as symmetric form of ${\rm
GL}(NM)$ $R$-matrix in the sense that the Planck constant and the
spectral parameter enter (almost) symmetrically. Such type
(symmetric) $R$-matrices are also shown to satisfy the Yang-Baxter
like quadratic and cubic equations.
 \end{abstract}

\newpage

{\small{

\tableofcontents

}}


\section{Introduction}
\setcounter{equation}{0}

\quad {\em The associative Yang-Baxter equation}
  \beq\label{z01}
  \begin{array}{c}
  \displaystyle{
 R_{12}
 R_{23}=R_{13}R_{12}+R_{23}R_{13}
 }
 \end{array}
 \eq
appeared in \cite{Aguiar} as coassociativity condition
$(\Delta\otimes {\rm id}_A)\circ\Delta=({\rm id}_A\otimes
\Delta)\circ\Delta$ for the (principle) derivation $\Delta:
A\rightarrow A\otimes A$ defined in an associative algebra $A$ by
$\Delta(a)=aR-Ra$, $\forall a\in A$ and some $R\in A\otimes A$. In
this paper $A=\Mat$, i.e. $N\times N$ matrices over $\mathbb C$.

The equation (\ref{z01}) was extended for $R$ depending on
additional parameters 
\cite{Pol} and used for
description of structures behind the classical Yang-Baxter equation
on elliptic curves and their degenerations \cite{Burban}. Let us
write it as
  \beq\label{x03}
  \begin{array}{c}
  \displaystyle{
 R^\hbar_{ab}
 R^{\eta}_{bc}=R^{\eta}_{ac}R_{ab}^{\hbar-\eta}+R^{\eta-\hbar}_{bc}R^\hbar_{ac}\,,\
 \ R^\hbar_{ab}=R^\hbar_{ab}(z_a\!-\!z_b)\,,
 }
 \end{array}
 \eq
(see (\ref{a190})) where $z_a,z_b,z_c,\hbar,\eta$ denote free
generic complex parameters. Indices $a,b,c$ are distinct numbers of
tensor components in $\Mat^{\otimes n\geq 3}$. $R_{ab}^\hbar$ is
defined in $a$-th and $b$-th components. If $c\neq a,b$ then
$R^\hbar_{ab}$ acts on the $c$-th component of $\Mat^{\otimes n}$ by
$N\!\times\! N$ identity matrix.

The elliptic solution of (\ref{x03}) was found in \cite{Pol}. It is
the Baxter-Belavin's elliptic ${\rm GL}_N$ quantum $R$-matrix
\cite{Baxter,Belavin} (in vector representation). The latter means
that $R^\hbar_{ab}$ satisfies the quantum Yang-Baxter
equation\footnote{See \cite{Baxter2} for reviews on the Yang-Baxter
equation and related structures.}:
  \beq\label{yb}
  \begin{array}{c}
  \displaystyle{
 R^\hbar_{ab}R^\hbar_{ac}R^\hbar_{bc}=R^\hbar_{bc}R^\hbar_{ac}R^\hbar_{ab}\,.
 }
 \end{array}
 \eq
%
%
%
%
 The property (\ref{x03}) of the Baxter-Belavin $R$-matrix was rediscovered in our papers
\cite{LOZ9,LOZ10} in the framework of integrable systems and related
topics -- Painlev\'e equations, Schlesinger systems and their
quantization via the Knizhnik-Zamolodchikov-Bernard (KZB) equations.

{\em Elliptic function identities.} 
The associative Yang-Baxter equation (\ref{x03}) in the scalar case
(when $N=1$ and $a,b,c=1,2,3$) is the Fay trisecant identity on
elliptic curve
  \beq\label{x02}
  \begin{array}{c}
  \displaystyle{
 \phi(z_1\!-\!z_2,\hbar)\phi(z_2\!-\!z_3,\eta)=\phi(z_1\!-\!z_3,\eta)\phi(z_1\!-\!z_2,\hbar\!-\!\eta)
 +\phi(z_2\!-\!z_3,\eta\!-\!\hbar)\phi(z_1\!-\!z_3,\hbar)
 }
 \end{array}
 \eq
The function $\phi(z,u)$ satisfying this equation is the Kronecker
function. It is define as
 \beq\label{x01}
 \begin{array}{c}
  \displaystyle{
 \phi(z,u)=\frac{\vth'(0)\vth(z+u)}{\vth(z)\vth(u)}\,,
  }
 \end{array}
 \eq
where $\vth(z)$ is the odd Riemann theta-function\footnote{In
trigonometric (hyperbolic) or rational cases
$\phi(z,u)=\coth(z)+\coth(u)$  or $\phi(z,u)=1/z+1/u$ respectively.}
(\ref{aa903}). In this respect the Baxter-Belavin's $R$-matrix
 can be viewed as non-abelian generalization of the
Kronecker function.  In \cite{LOZ10} we described a list of elliptic
function identities and properties together with their $R$-matrix
analogues.
 In particular, it is easy to verify that the Baxter-Belavin $R$-matrix defined
 as\footnote{In \cite{Pol} this $R$-matrix was found in different --
Richey-Tracy form \cite{RicheyT}.}
 \beq\label{x06}
 \begin{array}{c}
  \displaystyle{
R^\hbar_{12}(z)=\sum\limits_{\al\in\, {\mathbb Z}_N\times {\mathbb
Z}_N}\vf_\al(z,\om_\al+\hbar)\,T_\al\otimes T_{-\al}\,,\ \ \
\res\limits_{z=0}R^{\hbar}_{12}(z)=N P_{12}\,,
 }
 \end{array}
 \eq
(see definition of $\vf_\al(z,\om_\al+\hbar)$ in (\ref{a19}))
satisfies the skew-symmetry property
 \beq\label{x04}
 \begin{array}{c}
  \displaystyle{
R^\hbar_{ab}=-R^{-\hbar}_{ba}
 }
 \end{array}
 \eq
and the unitarity condition (see Appendix)
%
  \beq\label{x05}
  \begin{array}{c}
  \displaystyle{
 R^\hbar_{ab} R^{\hbar}_{ba}=1\otimes 1\,N^2(\wp(N\hbar)-\wp(z_a\!-\!z_b))\,,
 }
 \end{array}
 \eq
where $\wp(z)$ is the Weierstrass $\wp$-function. Condition
(\ref{x04}) is analogue of $\phi(z,u)=-\phi(-z,-u)$ while
(\ref{x05}) is similar to (\ref{aa906})
$\phi(z,u)\phi(z,-u)=\wp(z)-\wp(u)$.

In trigonometric (hyperbolic) or rational cases the Weierstrass
$\wp$-function entering (\ref{x05}) equals $\wp(z)=1/\sinh^2(z)$ or
$\wp(z)=1/z^2$ respectively. The simplest example of the $R$-matrix
satisfying (\ref{x03}), (\ref{yb}) and (\ref{x04}),(\ref{x05}) is
the quantum Yang's $R$-matrix \cite{Yang}:
  \beq\label{x07}
  \begin{array}{c}
  \displaystyle{
 R^{\rm Yang}_{12}(z)=\frac{1\otimes
1}{\hbar}+\frac{NP_{12}}{z}\,.
 }
 \end{array}
 \eq
It is the $R$-matrix analogue of function $1/\hbar+1/z$.
The class of $R$-matrices under consideration includes nontrivial
trigonometric and rational degenerations (of (\ref{x06})), which
satisfy (\ref{x03}),(\ref{x04}) and (\ref{x05}). These type
$R$-matrices can be found in \cite{Zabr} and \cite{Smirnov,LOZ8} for
trigonometric and rational cases respectively. In ${\rm GL}_2$ case
the corresponding 7-vertex and 11-vertex $R$-matrices were found in
\cite{Cherednik}. See also \cite{Burban} which results are
(presumably) gauge equivalent to those of \cite{Zabr,Smirnov,LOZ8}.

\vskip2mm

{\em Purpose of the paper.}

\vskip2mm

\noindent {\bf 1. Yang-Baxter equation with two Planck constants.}
As we will see it follows from (\ref{x03}), (\ref{x04}) and
(\ref{x05}) that the quantum $R$-matrix satisfies also cubic
equations of Yang-Baxter type. The one closest to the original one
(\ref{yb}) have the following form:
  \beq\label{x240}
  \begin{array}{c}
  \displaystyle{
 R^\eta_{12} R^\hbar_{13} R^\eta_{23}+R^\hbar_{12} R^\eta_{13} R^\hbar_{23}=R^\eta_{23}
 R^\hbar_{13} R^\eta_{12}+R^\hbar_{23}
 R^\eta_{13} R^\hbar_{12}\,,
 }
 \end{array}
 \eq
where again $R_{ab}^\hbar=R_{ab}^\hbar(z_a-z_b)$. When $\hbar=\eta$
(\ref{x240}) coincides with the Yang-Baxter
 equation (\ref{yb}).
Another (more general) type of equation contains $\wp$-function
entering the unitarity condition (\ref{x05}):
%
  \beq\label{x230}
  \begin{array}{c}
  \displaystyle{
 R^\eta_{12} R^\hbar_{13} R^\eta_{23}-R^\hbar_{23}
 R^\eta_{13} R^\hbar_{12}=R_{13}^{\hbar+\eta}\,N^2\left( \wp(N\eta)-\wp(N\hbar)
 \right)\,.
 }
 \end{array}
 \eq
Let us mention that equations of (\ref{x230}) type were considered
in \cite{Rub} in the context of double Lie (and the quadratic
Poisson) structures.

\vskip2mm

\noindent {\bf 2. Quantum algebras.} Given a quantum $R$-matrix one
can define the following (Sklyanin type) algebra using exchange
relations:
  \beq\label{x075}
  \begin{array}{c}
  \displaystyle{
R_{12}^\hbar(z-w)\hL_1^\hbar(z,\hS)\hL_2^\hbar(w,\hS)=\hL_2^\hbar(w,\hS)\hL_1^\hbar(z,\hS)R_{12}^\hbar(z-w)
 }
 \end{array}
 \eq
with ${\hat L}^\hbar(z,{\hat S})=\tr_2\left(R_{12}^\hbar(z)
\hS_2\right)$. For $N=2$, i.e. in the Baxter's case equation
(\ref{x075}) provides the Sklyanin algebra \cite{Sklyanin} for the
components $\hS_\al$, $\hS=\sum\limits_\al \hS_\al T_\al$. In
particular case (see (\ref{x53})) such $L$-operator is just the
$R$-matrix (\ref{x06}) itself, and the exchange relations
(\ref{x075}) are fulfilled identically due to Yang-Baxter equation
(\ref{yb}). It means that (\ref{x075}) follows from (\ref{yb}) by
treating one of tensor components as some Sklyanin algebra module.

In the same way we can define exchange relations corresponding to
(\ref{x240}) as follows:
  \beq\label{x076}
  \begin{array}{c}
  \displaystyle{
R_{12}^\hbar(z\!-\!w)\hL^\eta_1(z,\hS^\eta)
\hL^\hbar_2(w,\hS^\hbar)+R_{12}^\eta(z\!-\!w)\hL^\hbar_1(z,\hS^\hbar)
\hL^\eta_2(w,\hS^\eta)=\qquad \qquad \qquad \qquad}
 \\ \ \\
   \displaystyle{
 \qquad \qquad \qquad \qquad =\hL_2^\hbar(w,\hS^\hbar) \hL_1^\eta(z,\hS^\eta)
R_{12}^\hbar(z\!-\!w)+\hL_2^\eta(w,\hS^\eta)
\hL_1^\hbar(z,\hS^\hbar) R_{12}^\eta(z\!-\!w)
 }
 \end{array}
 \eq
Again, when $\hbar=\eta$ it coincides with the ordinary exchange
relations (\ref{x075}). As we will see this equation defines coupled
Sklyanin algebras, i.e. given a pair of Sklyanin algebras for
$\hS^\hbar$ and $\hS^\eta$ the equation (\ref{x076}) provides
commutation relations between $\hS^\hbar$ and $\hS^\eta$.

\vskip2mm

\noindent {\bf 3. Symmetric $R$-matrix in $\Mat^{\otimes 2}\otimes
\Matm^{\otimes 2}$.}
Here we consider the following extension of the Baxter-Belavin
$R$-matrix (\ref{x06}):
  \beq\label{x081}
  \begin{array}{c}
  \displaystyle{
\mR_{12,\ti 1\ti 2}(z,\hbar)=\sum\limits_{\ti\al\in\, {\mathbb
Z}_M\times {\mathbb Z}_M}\exp(2\pi\imath N\frac{\ti\al_2}{M}\hbar)\,
R_{12}^{\,\hbar}(z+N\om_{\ti\al})\otimes {\ti T}_{\ti\al}\otimes
{\ti T}_{-\ti\al}\,,
 }
 \end{array}
 \eq
 where ${\ti T}_{\ti\al}$ is $T_\al$ of size $M\times M$.
We call this $R$-matrix symmetric in the sense that both arguments
(the Planck constant $\hbar$ and the spectral parameter $z$) are
averaged over lattices ${\mathbb Z}_N\times {\mathbb Z}_N$ and
${\mathbb Z}_M\times {\mathbb Z}_M$ respectively. In the simplest
rational case corresponding to Yang's $R$-matrix (\ref{x07}) the
expression (\ref{x081}) is reduced  to
  \beq\label{x082}
  \begin{array}{c}
  \displaystyle{
 \mR_{12,\ti 1\ti 2}(z,\hbar)=M\,\frac{1_{N}\otimes
1_{N}\otimes {\ti P}_{\ti 1\ti 2}}{\hbar}+N\,\frac{P_{12}\otimes
{\ti 1}_M\otimes {\ti 1}_M}{z}\,.
 }
 \end{array}
 \eq
The limiting cases $N=1$ or $M=1$ are given as follows:
  \beq\label{x08}
  \begin{array}{ccc}
    &  \ \mR_{12,\ti 1\ti 2}(z,\hbar) &
 \\
   ^{M=1} \swarrow& &   \searrow^{N=1}
 \\
 R_{12}^{\,\hbar}(z) &    & R_{\ti 1 \ti 2}^{\,z}(\hbar)
 \end{array}
 \eq
We will show that such an $R$-matrix satisfies a set of relations
similar to those for the ordinary $R$-matrix. In particular, we have
the following generalization of the associative Yang-Baxter equation
(\ref{x02}):
  \beq\label{x083}
  \begin{array}{c}
  \displaystyle{
\mR_{12,\ti 1\ti 2}\, \mR_{23,\ti 3\ti 2}=\mR_{13,\ti 3\ti 2}\,
\mR_{12,\ti 1\ti 3} + \mR_{23,\ti 3\ti 1}\,\mR_{13,\ti 1\ti 2}\,,
 }
 \end{array}
 \eq
 where $\mR_{ab,\ti a\ti b}=\mR_{ab,\ti a\ti b}(z_{a}-z_b,\hbar_{\ti a}-\hbar_{\ti b})$.
The cubic equations arise similarly. For example, the
generalizations of (\ref{x230}) takes the form:
  \beq\label{x084}
  \begin{array}{l}
  \displaystyle{
\mR_{12,\ti 3\ti 2}\, \mR_{13,\ti 1\ti 3}\,\mR_{23,\ti 3\ti
2}=\mR_{23,\ti 1\ti 3}\,\mR_{13,\ti 3\ti 2}\, \mR_{12,\ti 1\ti
3}+N^2M^2\,\mR_{13,\ti 1\ti 2}\Big( \wp(N\hbar_{\ti 3\ti 2}) -
\wp(N\hbar_{\ti 1\ti 3}) \Big)\,,
 }
 \\ \ \\
  \displaystyle{
\mR_{32,\ti 1\ti 2}\, \mR_{13,\ti 1\ti 3}\,\mR_{32,\ti 2\ti
3}=\mR_{13,\ti 2\ti 3}\,\mR_{32,\ti 1\ti 3}\, \mR_{13,\ti 1\ti
2}+N^2M^2\,\mR_{12,\ti 1\ti 3}\Big( \wp(Mz_{23}) - \wp(Mz_{13})
\Big)\,.
 }
 \end{array}
 \eq
where $z_{ab}=z_a-z_b$ and $\hbar_{\ti a\ti b}=\hbar_{\ti
a}-\hbar_{\ti b}$. The Yang-Baxter like equation (\ref{x240}) is
generalized as follows:
  $$
  \begin{array}{l}
  \displaystyle{
\mR_{12,\ti 3\ti 2}(z_{12},\hbar_{\ti 3\ti 2})\, \mR_{13,\ti 1\ti
3}(z_{13},\hbar_{\ti 1\ti 3})\,\mR_{23,\ti 3\ti 2}(z_{23},\hbar_{\ti
3\ti 2})+\mR_{12,\ti 3\ti 2}(z_{12},\hbar_{\ti 1\ti 3})\,
\mR_{13,\ti 1\ti 3}(z_{13},\hbar_{\ti 3\ti 2})\,\mR_{23,\ti 3\ti
2}(z_{23},\hbar_{\ti 1\ti 3})
 }
 \\ \ \\
  \displaystyle{
=\mR_{23,\ti 1\ti 3}(z_{23},\hbar_{\ti 1\ti 3})\mR_{13,\ti 3\ti
2}(z_{13},\hbar_{\ti 3\ti 2}) \mR_{12,\ti 1\ti 3}(z_{12},\hbar_{\ti
1\ti 3})+\mR_{23,\ti 1\ti 3}(z_{23},\hbar_{\ti 3\ti 2})\mR_{13,\ti
3\ti 2}(z_{13},\hbar_{\ti 1\ti 3}) \mR_{12,\ti 1\ti
3}(z_{12},\hbar_{\ti 3\ti 2})
 }
 \end{array}
 $$
and
 $$
  \begin{array}{l}
  \displaystyle{
\mR_{32,\ti 1\ti 2}(z_{32},\hbar_{\ti 1\ti 2})\, \mR_{13,\ti 1\ti
3}(z_{13},\hbar_{\ti 1\ti 3})\,\mR_{32,\ti 2\ti 3}(z_{32},\hbar_{\ti
2\ti 3})+\mR_{32,\ti 1\ti 2}(z_{13},\hbar_{\ti 1\ti 2})\,
\mR_{32,\ti 1\ti 3}(z_{13},\hbar_{\ti 1\ti 3})\,\mR_{32,\ti 2\ti
3}(z_{13},\hbar_{\ti 2\ti 3})
 }
 \\ \ \\
  \displaystyle{
=\mR_{13,\ti 2\ti 3}(z_{13},\hbar_{\ti 2\ti 3})\mR_{32,\ti 1\ti
3}(z_{32},\hbar_{\ti 1\ti 3}) \mR_{13,\ti 1\ti 2}(z_{13},\hbar_{\ti
1\ti 2})+\mR_{13,\ti 2\ti 3}(z_{32},\hbar_{\ti 2\ti 3})\mR_{32,\ti
1\ti 3}(z_{13},\hbar_{\ti 1\ti 3}) \mR_{13,\ti 1\ti
2}(z_{32},\hbar_{\ti 1\ti 2})
 }
 \end{array}
 $$

\vskip3mm

{\small

\noindent {\bf Acknowledgments.} The work was supported by RFBR
grant 15-01-04217
and by joint RFBR project 15-51-52031 HHC$_a$. The work of A.L. was
partially supported by Department of Mathematics NRU HSE, the
subsidy granted to the HSE by the Government of the Russian
Federation for the implementation of the Global Competitiveness
Program and by the Simons Foundation. The work of A.Z. was also
partially supported by the D. Zimin's fund "Dynasty".

}

\section{Yang-Baxter equations with two Planck constants}
\setcounter{equation}{0}

Consider a quantum $R$-matrix which satisfies the associative
Yang-Baxter equation (\ref{x03}) and the properties (\ref{x04}),
(\ref{x05}). Then it is easy to get a set of cubic (in $R$)
relations.
 \begin{predl}
The following cubic relations are valid for a common solution of
(\ref{x03}), (\ref{x04}), (\ref{x05}):
  \beq\label{x21}
  \begin{array}{c}
  \displaystyle{
 R^\hbar_{12} R^\hbar_{13} R^\hbar_{23}=R^\hbar_{23}
 R^\hbar_{13} R^\hbar_{12}\,,
 }
 \end{array}
 \eq
  \beq\label{x22}
  \begin{array}{c}
  \displaystyle{
 R^\hbar_{ab} R^\hbar_{bc} R^\hbar_{ca}+R^\hbar_{ac}
 R^\hbar_{cb} R^\hbar_{ba}=-N^3\wp'(N\hbar)\,1_a\otimes 1_b\otimes 1_c\,,
 }
 \end{array}
 \eq
  \beq\label{x23}
  \begin{array}{c}
  \displaystyle{
 R^\eta_{ab} R^\hbar_{ac} R^\eta_{bc}-R^\hbar_{bc}
 R^\eta_{ac} R^\hbar_{ab}=R_{ac}^{\hbar+\eta}\,N^2\left( \wp(N\eta)-\wp(N\hbar) \right)\,,
 }
 \end{array}
 \eq
  \beq\label{x24}
  \begin{array}{c}
  \displaystyle{
 R^\eta_{ab} R^\hbar_{ac} R^\eta_{bc}+R^\hbar_{ab} R^\eta_{ac} R^\hbar_{bc}=R^\eta_{bc}
 R^\hbar_{ac} R^\eta_{ab}+R^\hbar_{bc}
 R^\eta_{ac} R^\hbar_{ab}\,,
 }
 \end{array}
 \eq
where $R_{ab}^\hbar=R^\hbar_{ab}(z_a-z_b)$ and $a,b,c$ are distinct
numbers from the set $\{1,2,3\}$.
 \end{predl}
\noindent\underline{\emph{Proof:}}  First, notice that the primary
one is the third identity (\ref{x23}) since all others follow from
it. Indeed, the Yang-Baxter equation (\ref{x21}) follows from
(\ref{x23}) in the case $\hbar=\eta$.

The second equation (\ref{x22}) appears from (\ref{x23}) in the
limit $\eta\rightarrow -\hbar\,$: one should use the classical limit
(\ref{x051}), and then the skew-symmetry property (\ref{x04}) to
arrange the indices in a cyclic order. This gives (\ref{x22}) up to
permutation of indices.

The fourth relation (\ref{x24}) can be called Yang-Baxter equation
with two Planck constants because its structure is similar to
(\ref{x21}) and it is coincide with (\ref{x21}) for $\hbar=\eta$. It
is easy to see that (\ref{x24}) follows from (\ref{x23}) as skew
symmetry of (\ref{x23}) l.h.s. with respect to interchanging $\hbar$
and $\eta$.

Thus we need to prove  (\ref{x23}). Consider (\ref{x03}) and
multiply both parts by $R_{bc}^{\hbar-\eta}$ from the left side:
  \beq\label{x25}
  \begin{array}{c}
  \displaystyle{
 R_{bc}^{\hbar-\eta}R^\hbar_{ab}
 R^{\eta}_{bc}=R_{bc}^{\hbar-\eta}R^{\eta}_{ac}R_{ab}^{\hbar-\eta}+R_{bc}^{\hbar-\eta}R^{\eta-\hbar}_{bc}R^\hbar_{ac}\,.
 }
 \end{array}
 \eq
Now interchange the indices 2,3 in (\ref{x03}) and multiply both
parts by $R_{bc}^\eta$
  \beq\label{x26}
  \begin{array}{c}
  \displaystyle{
 R^\hbar_{ac}
 R^{\eta}_{cb}R_{bc}^\eta=R^{\eta}_{ab}R_{ac}^{\hbar-\eta}R_{bc}^\eta-R^{\hbar-\eta}_{bc}R^\hbar_{ab}R_{bc}^\eta\,,
 }
 \end{array}
 \eq
where in the last term we have already used
$R^{\eta-\hbar}_{cb}=-R^{\hbar-\eta}_{bc}$. From (\ref{x04}),
(\ref{x05}) it follows that $R_{bc}^{\hbar-\eta}R^{\eta-\hbar}_{bc}$
and $R^{\eta}_{cb}R_{bc}^\eta$ are scalar operators. Subtracting
(\ref{x26}) from (\ref{x25}) we get (\ref{x23}) with
$\hbar:=\hbar-\eta$. $\blacksquare$

Let us also remark that the identity (\ref{x22}) appeared to be
related to $R$-matrix valued Lax operator for the classical
Calogero-Moser model \cite{LOZ9}. Consider the following
block-matrix:
  \beq\label{x27}
 \begin{array}{c}
  \displaystyle{
\mathcal L(\hbar)=\sum\limits_{a,b=1}^{n} \ti{ E}_{ab}\otimes
(1-\delta_{ab})R_{ab}^\hbar(z_a-z_b)\,,
 }
 \end{array}
 \eq
where $\ti{E}_{ab}$ is the standard basis in ${\rm Mat}(n,\mathbb
C)$, i.e. $(\ti{ E}_{ab})_{cd}=\delta_{ac}\delta_{bd}$. Then the
diagonal blocks of $\tr \mathcal L^k(\hbar)$ are scalar operators,
and the scalar functions are obtained as if $\mathcal L(\hbar)$ were
element of ${\rm Mat}(n,\mathbb C)$ with matrix elements
$l_{ab}(\hbar)=(1-\delta_{ab})N\phi(N\hbar,z_a-z_b)$:
  \beq\label{x28}
 \begin{array}{c}
  \displaystyle{
\left(\mathcal L^k(\hbar)\right)_{aa}=1\otimes...\otimes 1 \left(
l^k(\hbar)\right)_{aa}\,.
 }
 \end{array}
 \eq
When $n=2$ this equation is equivalent to the unitarity condition
(\ref{x05}), while for $n=3$ (\ref{x28}) reproduces (\ref{x22}).

One more application of (\ref{x22}) comes from the classical limit
(\ref{x051}). The identities which appear from (\ref{x22}) together
with (\ref{x055}) provide sufficient conditions for compatibility of
the KZB connections (see details in \cite{LOZ9}).

Finally we conclude that the identity (\ref{x23}) is of great
importance. On one hand it reproduces the Yang-Baxter equations, and
in this sense it "knows" about quantum integrability and related
algebraic structures including quantum groups, Sklyanin algebras,
e.t.c. On the other hand (\ref{x23}) also "knows" about classical
integrable system of Calogero type. At last, the same identity
provides the quantization of the Schlesinger systems via
compatibility of the KZB connections.

\section{Quantum algebras}
\setcounter{equation}{0}

In this section we discuss differen types of quadratic
finite-dimensional quantum algebras arising from exchange like
relations. The quantum (Lax) $L$-operator \cite{Sklyanin,Hasegawa}
is defined as
  \beq\label{x51}
  \begin{array}{c}
  \displaystyle{
 {\hat L}^\hbar(z,\hS)={\hat L}^\hbar(z)=\tr_2(R^\hbar_{12}(z) {\hat S}_2)\,, \ \ \
 \hS=\sum\limits_\al T_\al \hS_\al
 }
 \end{array}
 \eq
where $\{\hS_\al,\ \al\in{\mathbb Z}_N^{\times 2}\}$ is the set of
generators of the quantum algebra $\mathcal A$. For the elliptic
$R$-matrix (\ref{x06}) we have
  \beq\label{x52}
  \begin{array}{c}
  \displaystyle{
 {\hat L}^\hbar(z)=\sum\limits_\al T_\al \hS_\al
 \vf_\al(z,\om_\al+\hbar)\,.
 }
 \end{array}
 \eq
For the vector representation $\rho_N$ of $\mathcal A$ given by
  \beq\label{x53}
  \begin{array}{c}
  \displaystyle{
 \rho_N(\hS_\al)=T_{-\al}\in\Mat
 }
 \end{array}
 \eq
the Lax operator (\ref{x52}) coincides with the $R$-matrix
(\ref{x06}).

\noindent {\bf Finite Heisenberg group} (see (\ref{a17}) in
Appendix)
is the simplest example of quantum algebra which comes from
$R$-matrix relations for the $L$-operator (\ref{x51}).
\begin{predl}
 Relations
  \beq\label{x54}
  \begin{array}{c}
  \displaystyle{
\hL_1^\hbar(z) \hL_2^\eta(w)=\hL_2^{\hbar+\eta}(w)
R_{12}^\hbar(z-w)-R_{12}^{-\eta}(z-w) \hL_1^{\hbar+\eta}(z)
 }
 \end{array}
 \eq
 or
  \beq\label{x541}
  \begin{array}{c}
  \displaystyle{
\hL_2^\eta(w) \hL_1^\hbar(z) =
R_{12}^\hbar(z-w)\hL_2^{\hbar+\eta}(w)-\hL_1^{\hbar+\eta}(z)
R_{12}^{-\eta}(z-w)
 }
 \end{array}
 \eq
 with $L$-operator defined in (\ref{x52}) and $R$-matrix (\ref{x06})
 are equivalent to
 \beq\label{x55}
  \begin{array}{c}
  \displaystyle{
 \hS_1\,\hS_2=N P_{12}\,\hS_1\,,
 }
 \end{array}
 \eq
i.e.
 \beq\label{x56}
  \begin{array}{c}
  \displaystyle{
\hS_\al\,\hS_\be=\kappa_{\al,\be}\,\hS_{\al+\be}
 }
 \end{array}
 \eq
in components. The associativity condition for the triple product
$\hL_1^\hbar(z)\hL_2^\eta(w)\hL_3^\xi(x)$ follows from the
associative Yang-Baxter equation (\ref{x03}).
\end{predl}
The proof of (\ref{x54}), (\ref{x541}) follows directly from the Fay
identity (\ref{x02}). The statement about associativity is analogues
to the derivation of (\ref{z01}) in \cite{Aguiar}.

Notice that the commutation relations
  \beq\label{x57}
  \begin{array}{c}
  \displaystyle{
[\hL_1^\hbar(z), \hL_2^\eta(w)]=[\hL_1^{\hbar+\eta}(z),
R_{12}^{-\eta}(z-w)]+[\hL_2^{\hbar+\eta}(w), R_{12}^\hbar(z-w)]
 }
 \end{array}
 \eq
are equivalent to those for the Lie algebra ${\rm gl}_N$:
  \beq\label{x571}
  \begin{array}{c}
  \displaystyle{
[\hS_1,\hS_2]=N[P_{12},\hS_1]\,.
 }
 \end{array}
 \eq
Obviously the same expression (r.h.s. of (\ref{x57})) can be used
for definition of the Poisson-Lie brackets
$\{S_1,S_2\}=N[P_{12},S_1]$ on ${\rm gl}_N^*$:
  \beq\label{x572}
  \begin{array}{c}
  \displaystyle{
\{L_1^\hbar(z), L_2^\eta(w)\}=[L_1^{\hbar+\eta}(z),
R_{12}^{-\eta}(z-w)]+[L_2^{\hbar+\eta}(w), R_{12}^\hbar(z-w)]\,.
 }
 \end{array}
 \eq
 with the classical matrix-valued function $L(z,S)$ on the phase
 space ${\rm gl}_N^*$
 parameterized by $S_\al$ ($S=\sum_\al S_\al T_\al$). The Jacobi
 identity is due to (\ref{x03}).
The relation (\ref{x572}) differs from the custom classical exchange
relations
  \beq\label{x573}
  \begin{array}{c}
  \displaystyle{
\{l_1(z), l_2(w)\}=[l_1(z), r_{12}(z-w)]+[l_2(w), r_{12}(z-w)]\,,
 }
 \end{array}
 \eq
 where the classical $r$-matrix $r_{12}(z)$ (\ref{x051}) is
used, and the Jacobi identity is fulfilled due the classical
Yang-Baxter equation (\ref{x052}) for $r_{12}(z)$.
Alternatively, we can say that in the case under consideration the
constant $\hbar$ (and $\eta$) is not the Planck constant entering
the quantum $R$-matrix but rather additional spectral parameter
entering the classical $r$-matrix. Indeed, the final commutation
relations (\ref{x571}) are independent of $\hbar,\eta$ (as well as
they are independent of $z$ and $w$). Such interpretation is close
to the consideration suggested in \cite{Pol}.

\noindent {\bf Exchange relations with two Planck constants and
coupled Sklyanin algebras.} Consider a pair of the quantum Lax
operators
  \beq\label{x5780}
  \begin{array}{c}
  \displaystyle{
  {\hat L}^\hbar(z)={\hat L}^\hbar(z,{\hat S}^\hbar)=\sum\limits_\al T_\al \hS_\al^\hbar
 \vf_\al(z,\om_\al+\hbar)
 }
 \end{array}
 \eq
and
  \beq\label{x57801}
  \begin{array}{c}
  \displaystyle{
  {\hat L}^\eta(z)={\hat L}^\eta(z,{\hat S}^\eta)=\sum\limits_\al T_\al \hS_\al^\eta
 \vf_\al(z,\om_\al+\eta)
 }
 \end{array}
 \eq
Each of them defines its own Sklyanin algebra via exchange relations
(\ref{x075}). See details in the Appendix. Here we suggest another
exchange type relation which provides commutation relations between
$\hS^\hbar$ and $\hS^\eta$. It is of the form:
  \beq\label{x578}
  \begin{array}{c}
  \displaystyle{
R_{12}^\hbar(z\!-\!w)\hL^\eta_1(z)
\hL^\hbar_2(w)+R_{12}^\eta(z\!-\!w)\hL^\hbar_1(z)
\hL^\eta_2(w)=\qquad \qquad \qquad \qquad}
 \\ \ \\
   \displaystyle{
 \qquad \qquad \qquad \qquad =\hL_2^\hbar(w) \hL_1^\eta(z)
R_{12}^\hbar(z\!-\!w)+\hL_2^\eta(w) \hL_1^\hbar(z)
R_{12}^\eta(z\!-\!w)
 }
 \end{array}
 \eq

\begin{predl}
 Equation (\ref{x578}) for the Lax operators $\hL^\hbar(\hS^\hbar)$ and $\hL^\eta(\hS^\eta)$ is equivalent to the following commutation
 relations for the components of $\hS^\hbar$ and $\hS^\eta$:
  \beq\label{x5781}
  \begin{array}{c}
  \displaystyle{
 \sum\limits_\ga
 \kappa_{\ga,\al-\be}\hS_{\al-\ga}^\eta \hS_{\be+\ga}^\hbar\, {\rm f}^{\hbar,\eta}_{\al,\be,\ga}+
 \kappa_{\ga,\al-\be}\hS_{\al-\ga}^\hbar \hS_{\be+\ga}^\eta\, {\rm f}^{\eta,\hbar}_{\al,\be,\ga}
 =0\,,
 }
 \end{array}
 \eq
 where the structure constants ${\rm f}^{\hbar,\eta}_{\al,\be,\ga}$ are given by
 \beq\label{x5782}
 \begin{array}{l}
  \displaystyle{
{\rm for}\ \be\neq 0:\quad {\rm
f}^{\hbar,\eta}_{\al,\be,\ga}=E_1(\om_\ga+\hbar)-E_1(\om_{\al-\be-\ga}+\eta)+E_1(\om_{\al-\ga}+\eta)-E_1(\om_{\be+\ga}+\hbar)\,,
 }
 \\ \ \\
  \displaystyle{
{\rm for}\ \be=0:\quad {\rm
f}^{\hbar,\eta}_{\al,0,\ga}=\wp(\om_\ga+\hbar)-\wp(\om_{\al-\ga}+\eta)\,.
 }
 \end{array}
 \eq
\end{predl}
\noindent\underline{\emph{Proof:}} Here we use short notation
$\vf_\al^\hbar(z)=\vf_\al^\hbar(z,\om_\al+\hbar)$. Compute
 $$
 R_{12}^\hbar(z\!-\!w)\hL_1^\eta(z)\hL_2^\hbar(w)-\hL_2^\eta(w)\hL_1^\hbar(z)R_{12}^\eta(z\!-\!w)=
  \sum\limits_{\al,\be,\ga} T_\al\otimes T_\be\times
 $$
 $$
 \times\left(
\hS^\eta_{\al-\ga}\hS^\hbar_{\be+\ga} \kappa_{\ga,\al-\be}
\vf_{\al-\ga}^\eta(z) \vf_{\be+\ga}^\hbar(w) \vf_\ga^\hbar(z\!-\!w)
 -
\hS^\eta_{\be+\ga}\hS^\hbar_{\al-\ga} \kappa_{\al-\be,\ga}
\vf_{\al-\ga}^\hbar(z) \vf_{\be+\ga}^\eta(w) \vf_\ga^\eta(z\!-\!w)
\right)=
 $$
By redefining the summation index for the second term as
$\ga\rightarrow \al-\be-\ga$  we get for the following coefficient
behind the tensor component $T_\al\otimes T_\be$:
 $$
=\sum\limits_{\al,\be,\ga} T_\al\otimes
T_\be\,\hS^\eta_{\al-\ga}\hS^\hbar_{\be+\ga} \kappa_{\ga,\al-\be}
\left( \vf_{\al-\ga}^\eta(z) \vf_{\be+\ga}^\hbar(w)
\vf_\ga^\hbar(z\!-\!w)
-\vf_{\be+\ga}^\hbar(z)\vf_{\al-\ga}^\eta(w)\vf_{\al-\be-\ga}^\eta(z\!-\!w)
\right)
 $$
 $$
=\sum\limits_{\al,\be,\ga} T_\al\otimes
T_\be\,\hS^\eta_{\al-\ga}\hS^\hbar_{\be+\ga}\,
\kappa_{\ga,\al-\be}\, {\rm
f}^{\hbar,\eta}_{\al,\be,\ga}\,\vf_\al^{\hbar+\eta}(z) \vf_\be(w)\,,
 $$
where we assume $\vf_0(w)=1$. The last equality follows from
(\ref{aa907}), (\ref{aa908}). In this way we reproduce the first
term in (\ref{x5781}). The second term in (\ref{x5781}) is obtained
in the same way by interchanging $\hbar$ and $\eta$. $\blacksquare$

Notice that when $\hbar=\eta$ the structure constants ${\rm
f}^{\hbar,\hbar}_{\al,\be,\ga}={\rm f}^{\hbar}_{\al,\be,\ga}$
coincide with those for the Sklyanin algebra (\ref{x62}).


Recently the coupled Sklyanin algebras appeared also in different
way -- via the modular double \cite{Spiridonov}. In that case two
algebras depend on different modular parameters $\tau$. In our case
this parameter is the same but the Planck constants are different.

{\bf Remark.} One can also consider (quasi)classical limits of the
Yang-Baxter (\ref{x24}) and exchange relations (\ref{x578}). There
are different possibilities for the limits since we deal with two
Planck constants. In particular, when $\hbar\rightarrow 0$ (while
$\eta$ is finite) (\ref{x24}) is reduced to
  \beq\label{x99}
  \begin{array}{c}
  \displaystyle{
[R^\eta_{12},R^\eta_{23}]+[R_{13}^\eta,r_{23}]+[r_{12},R^\eta_{13}]=0\,,
 }
 \end{array}
 \eq
 where $r_{ab}$ is the classical $r$-matrix (\ref{x051}). At this
 stage the generators $\hS^\hbar_\al$ become classical variables
 $S_\al$ while $\hS^\eta_\al$ are still quantum. Then we get a
 "half-quantum" Poisson structure between the commutative classical
 variables $S_\al$ and noncommutative variables $\hS^\eta_\be$.
 Taking then limit $\eta\rightarrow 0$ the equation (\ref{x99})
 is reduced to the ordinary classical Yang-Baxter equation (\ref{x052}),
 and the Poisson structure is the standard classical Sklyanin
 algebra. We will describe the quasiclassical limits in our next
 paper.

\section{Symmetric $R$-matrix in $\Mat^{\otimes 2}\otimes \Matm^{\otimes 2}$}
\setcounter{equation}{0}

In this section we consider an extension of the $R$-matrix
(\ref{x06}) in $\Mat^{\otimes 2}$ to the matrix function in
$\Mat^{\otimes 2}\otimes \Matm^{\otimes 2}$. This type of
$R$-matrices (see e.g. \cite{Maillet,LOSZ3,Avan}) appears naturally
when $N:=N\times M$ from the original $\Mat^{\otimes 2}$-valued one.
Our definition is slightly modified in a way to get expression which
is symmetric in spectral parameter $z$ and the Planck constant
$\hbar$. The simplest example of the $R$-matrix which we are going
to discuss here can be given in the rational case. Consider the
following expression:
  \beq\label{x74}
  \begin{array}{c}
  \displaystyle{
 \mR_{12,\ti 1\ti 2}(z_1-z_2,\hbar_{\ti 1}-\hbar_{\ti 2})=M\,\frac{1_{N}\otimes
1_{N}\otimes {\ti P}_{\ti 1\ti 2}}{\hbar_{\ti 1}-\hbar_{\ti
2}}+N\,\frac{P_{12}\otimes {\ti 1}_M\otimes {\ti 1}_M}{z_1-z_2}\,,
 }
\\ \ \\
   \displaystyle{
 \mR_{12,\ti 1\ti 2}(z_1-z_2,\hbar_{\ti 1}-\hbar_{\ti 2})\in
\Mat^{\otimes 2}\otimes \Matm^{\otimes 2}\,,
 }
 \end{array}
 \eq
 where ${\ti P}_{\ti 1\ti 2}$ is the permutation operator in $\Matm^{\otimes
 2}$, and $1_N$ (or ${\ti 1}_M$) is the identity matrix in ${\rm Mat}(N,\mathbb C)$ (or in ${\rm Mat}(M,\mathbb C)$).
$\mR_{12,\ti 1\ti 2}$ has four tensor indices. The first pair is for
the components of $\Mat^{\otimes 2}$, and the second pair is for the
components of $\Matm^{\otimes 2}$. When $M=1$ (\ref{x74}) coincides
with the Yang's $R$-matrix (\ref{x07}) $R_{12}^{\hbar_{\ti
1}-\hbar_{\ti 2}}(z_1-z_2)\in\Mat^{\otimes 2}$. In the same way when
$N=1$ (\ref{x74}) gives $R_{12}^{z_1-z_2}(\hbar_{\ti 1}-\hbar_{\ti
2})\in\Matm^{\otimes 2}$.

The expression (\ref{x74}) is not a new $R$-matrix of course.
Multiplying it by $1_{N}\otimes 1_{N}\otimes {\ti P}_{\ti 1\ti 2}$
we get a special form of the Yang's $R$-matrix in ${\rm
Mat}(NM,\mathbb C)^{\otimes 2}$.

Below we define elliptic analogue of (\ref{x74}). But first we
recall some important properties of the Baxter-Belavin's $R$-matrix
which will be to extended to the case $\Mat^{\otimes 2}\otimes
\Matm^{\otimes 2}$ for the proper definition of underlying set of
functions. To be exact, our goal is to define a set of functions
$\Phi_{\al,\ti\al}(z,\hbar)$, $\al\in {\mathbb Z}_N\times {\mathbb
Z}_N$, $\ti\al\in {\mathbb Z}_M\times {\mathbb Z}_M$  which are
invariant with respect to shift by the full lattice periods
$\al\rightarrow \al+N$, $\al\rightarrow \al+N\tau$ and
$\ti\al\rightarrow \ti\al+M$, $\ti\al\rightarrow \ti\al+M\tau$.

For the reasons given above (and for the brevity sake) we will refer
to $\mR_{12,\ti 1\ti 2}$ type expression as {\em symmetric}
$R$-matrix though it is not such symmetric in the elliptic case as
in (\ref{x74}).


\noindent{\bf Symmetries of Baxter-Belavin $R$-matrix.}
The quantum elliptic $R$-matrix (\ref{x06}) is ${\mathbb Z}_N\times
{\mathbb Z}_N$ symmetric, i.e.
  \beq\label{x7499}
  \begin{array}{c}
  \displaystyle{
R^\hbar_{12}(z)=g_1^{-1}g_2^{-1}\,R_{12}^\hbar(z)\,g_1\,
g_2\,,\qquad g_1=g\otimes 1\,,\ g_2=1\otimes g\,,
 }
 \end{array}
 \eq
 where $g=Q$ or $g=\Lambda$ (\ref{a14}).
This $R$-matrix is the quasiperiodic function of the spectral
parameter $z$
  \beq\label{x749}
  \begin{array}{c}
  \displaystyle{
R^\hbar_{12}(z+1)=Q_1^{-1}\,R_{12}^\hbar(z)\,Q_1\,,
 }
\\ \ \\
  \displaystyle{
R^\hbar_{12}(z+\tau)=\exp(-2\pi\imath\hbar)\,\Lambda_1^{-1}\,R_{12}^\hbar(z)\,\Lambda_1
 }
 \end{array}
 \eq
on the elliptic curve with periods $1,\tau$. At the same time it is
the quasiperiodic function of the Planck constant $\hbar$ on the
smaller elliptic curve with periods $1/N,\tau/N$:
  \beq\label{x750}
  \begin{array}{c}
  \displaystyle{
R^{\hbar+1/N}_{12}(z)=Q_1^{-1}\,R_{12}^\hbar(z)\,Q_2\,,
 }
\\ \ \\
  \displaystyle{
R^{\hbar+\tau/N}_{12}(z)=\exp(-2\pi\imath
z/N)\,\Lambda_1^{-1}\,R_{12}^\hbar(z)\,\Lambda_2\,.
 }
 \end{array}
 \eq
The property (\ref{x750}) can be considered as a consequence of
(\ref{x749}) and the arguments symmetry property
  \beq\label{x748}
  \begin{array}{l}
  \displaystyle{
R^\hbar_{12}(z)=R_{12}^{\frac{z}{N}}(N\hbar)\,P_{12}\,,
 }
 \end{array}
 \eq
 which follows from (\ref{aa909})-(\ref{a910}). In order to verify
 (\ref{x750}) directly consider
 $$
 (T_\ga)^{-1}_1 R_{12}^\hbar(z) (T_\ga)_2
 $$
for some fixed $T_\ga$ (\ref{a16}). Here the lower indices 1,2 are
the numbers of the tensor components (\ref{x7499}), and
$T_\ga^{-1}=T_{-\ga}$ due to (\ref{a17}). For the above expression
we have:
 $$
=\sum\limits_{\al\in\, {\mathbb Z}_N\times {\mathbb
Z}_N}\vf_\al(z,\om_\al+\hbar)\,T_{-\ga}T_\al\otimes
T_{-\al}T_{\ga}=\sum\limits_{\al\in\, {\mathbb Z}_N\times {\mathbb
Z}_N}\vf_\al(z,\om_\al+\hbar)\,T_{\al-\ga}\otimes T_{\ga-\al}
 $$
Now shift the summation indices $\al\rightarrow \al+\ga$. Then from
(\ref{a19}) we get
  \beq\label{x751}
  \begin{array}{c}
  \displaystyle{
 (T_\ga)^{-1}_1 R_{12}^\hbar(z) (T_\ga)_2=\exp(2\pi\imath
 \p_\tau\om_\ga) R^{\hbar+\om_\ga}_{12}(z)\,.
 }
 \end{array}
 \eq
This answer reproduces (\ref{x750}) for $\ga=(1,0)$ and $\ga=(0,1)$
(then $T_\ga=Q$, $\om_\ga=1/N$ and $T_\ga=\Lambda$, $\om_\ga=\tau/N$
respectively).

Notice that in the proof of (\ref{x750}) we used the shift of
summation indices ($\al\rightarrow \al+\ga$). At the same time the
limits of the sum were not changed because  $\al\in\, {\mathbb
Z}_N\!\times\! {\mathbb Z}_N$. In fact, we also used the periodicity
of functions (\ref{a912}) $\vf_\al(z,\om_\al+\hbar)$ in indices
considered as discrete variables:
  \beq\label{x752}
  \begin{array}{c}
  \displaystyle{
\vf_{\al_1+N,\al_2}(z,\om_\al+\om_{(N,0)}+\hbar)=\vf_{\al_1,\al_2+N}(z,\om_\al+\om_{(0,N)}+\hbar)=
\vf_\al(z,\om_\al+\hbar)\,.
 }
 \end{array}
 \eq
The property (\ref{x752}) demonstrates necessity of the exponential
factor $\exp(2\pi\imath\p_\tau\om_\al)$ in the definition
(\ref{a912}).

\vskip2mm

\noindent {\bf Symmetric $R$-matrix in $\Mat^{\otimes 2}\otimes
\Matm^{\otimes 2}$.} In this section we assume that $N$ and $M$ are
coprime integers unless otherwise specified\footnote{The general
case is not difficult but requires more additional notations. We
consider special case $M=N$ in the next section.}.

Consider the following expression:
 \beq\label{x71}
 \begin{array}{c}
  \displaystyle{
   \mR_{12,\ti 1\ti 2}(z,\hbar)
=\sum\limits_{\al\in\, {\mathbb Z}_N\times {\mathbb Z}_N}
   \sum\limits_{\ti\al\in\, {\mathbb Z}_M\times {\mathbb
   Z}_M} \Phi_{\al,\ti\al}(z,\hbar)\,T_\al\otimes T_{-\al}\otimes{\ti T}_{\ti\al}\otimes {\ti
   T}_{-\ti\al}\,,
   }
 \end{array}
 \eq
 where (with the definition (\ref{a912}))
 \beq\label{x72}
 \begin{array}{c}
  \displaystyle{
  \Phi_{\al,\ti\al}(z,\hbar)=
  \exp\left( 2\pi\imath(z+N{\ti\om}_{\ti\al})\frac{\al_2}{N} + 2\pi\imath\hbar\, N\frac{{\ti\al}_2}{M} \right)
   \phi(z+N{\ti\om}_{\ti\al}\,,\hbar+\om_\al)=
  }
 \end{array}
 \eq
 $$
=\exp\left( 2\pi\imath\hbar\, N\frac{{\ti\al}_2}{M} \right)
\vf_{\al}(z+N{\ti\om}_{\ti\al}\,,\hbar+\om_\al)
 $$
and for any $\al=(\al_1,\al_2)\in{\mathbb Z}_N\times {\mathbb Z}_N$
and $\ti\al=({\ti\al}_1,{\ti\al}_2)\in {\mathbb Z}_M\times {\mathbb
   Z}_M$
 \beq\label{x73}
 \begin{array}{c}
  \displaystyle{
  \om_\al=\frac{\al_1+\al_2\tau}{N}\,,\qquad {\ti\om}_{\ti\al}=\frac{ {\ti\al}_1+{\ti\al}_2\tau
  }{M}\,,
  }
 \end{array}
 \eq
while $\{T_\al\}$  (and $\{{\ti T}_{\ti\al}\}$) is the basis in
$\Mat$ (and $\Matm$ respectively) defined as in
(\ref{a14})-(\ref{a17}). That is, we mark the $\Matm$ elements and
related quantities (including related indices) by tildes. For
example, ${\ti\kappa}_{{\ti\al},{\ti\be}}=\exp\left( {\pi
\imath\,}({\ti\be}_1 {\ti\al}_2-{\ti\be}_2{\ti\al}_1)/M\right)$. For
coprime $N$ and $M$ two sets  $\{f(N\ti\om_{\ti\al})\,,\ \ti\al\in
{\mathbb Z}_M\times {\mathbb Z}_M\}$ and $\{f(\ti\om_{\ti\al})\,,\
\ti\al\in {\mathbb Z}_M\times {\mathbb Z}_M\}$ are equal for a
periodic function $f$ on the lattice ${\mathbb Z}_M\times {\mathbb
Z}_M$.

First, notice that for $M=1$ (then $\ti\al=(0,0)$) expression
(\ref{x71}) reproduces the definition of the Belavin's $\Mat$
$R$-matrix (\ref{x06}) $R^\hbar_{12}(z)\in\Mat^{\otimes 2}$, and
similarly for $N=1$ we have $R^z_{12}(\hbar)\in\Matm^{\otimes 2}$.

The analogues of the quasiperiodic properties  (boundary conditions)
on the lattice $\mathbb Z\oplus\tau\mathbb Z$ are of the form:
  \beq\label{x33}
  \begin{array}{c}
  \displaystyle{
 \mR_{12,\ti 1\ti 2}(z+1,\hbar)=Q_1^{-1}\,\mR_{12,\ti 1\ti
 2}(z,\hbar)\,
 Q_1\,,
 }
 \\ \ \\
  \displaystyle{
 \mR_{12,\ti 1\ti 2}(z+\tau,\hbar)=\exp(-2\pi\imath\hbar)\,\Lambda_1^{-1}\,\mR_{12,\ti 1\ti
 2}(z,\hbar)\,
 \Lambda_1
 }
 \end{array}
 \eq
and
  \beq\label{x34}
  \begin{array}{c}
  \displaystyle{
 \mR_{12,\ti 1\ti 2}(z,\hbar+1)={\ti Q}_{\ti 1}^{-N}\,\mR_{12,\ti 1\ti
 2}(z,\hbar)\,
 {\ti Q}_{\ti 1}^N\,,
 }
 \\ \ \\
  \displaystyle{
 \mR_{12,\ti 1\ti 2}(z,\hbar+\tau)=\exp(-2\pi\imath z)\,{\ti \Lambda}_{\ti 1}^{-N}\,\mR_{12,\ti 1\ti
 2}(z,\hbar)\,
 {\ti \Lambda}_{\ti 1}^N\,,
 }
 \end{array}
 \eq
where $Q_1=Q\otimes 1_N\otimes {\ti 1}_M\otimes {\ti 1}_M$, ${\ti
Q}_{\ti 1}=1_N\otimes 1_N\otimes {\ti Q}\otimes {\ti 1}_M$, and $\ti
Q$, $\ti\Lambda$ are the matrices of form (\ref{a14}) but of size
$M\times M$.

It is important for us that functions $\Phi_{\al,\ti\al}(z,\hbar)$
are periodic in both discrete variables -- indices
$\al=(\al_1,\al_2)\in {\mathbb Z}_N\times {\mathbb Z}_N$ and
$\ti\al=({\ti\al}_1,{\ti\al}_2)\in {\mathbb Z}_M\times {\mathbb
Z}_M$, i.e. similarly to (\ref{x752}) we have
  \beq\label{x342}
  \begin{array}{c}
  \displaystyle{
\Phi_{(\al_1+N,\al_2),\ti\al}(z,\hbar)=\Phi_{(\al_1,\al_2+N),\ti\al}(z,\hbar)=\Phi_{(\al_1,\al_2),\ti\al}(z,\hbar)
 }
 \end{array}
 \eq
and
  \beq\label{x343}
  \begin{array}{c}
  \displaystyle{
\Phi_{\al,({\ti\al}_1+M,{\ti\al}_2)}(z,\hbar)=\Phi_{\al,({\ti\al}_1,{\ti\al}_2+M)}(z,\hbar)
=\Phi_{\al,({\ti\al}_1,{\ti\al}_2)}(z,\hbar)\,.
 }
 \end{array}
 \eq
These properties allow to shift indices of summations in the same
way as it was used in (\ref{x751}). To verify
(\ref{x33})-(\ref{x34}) and (\ref{x342})-(\ref{x343}) one needs
(\ref{aa9031}).

In ${\rm GL}_N$ case we had also quasiperiodic behavior (\ref{x750})
on the smaller elliptic curve with periods $1/N$ and $\tau/N$. For
the symmetric $R$-matrix (\ref{x750}) acquires the form:
  \beq\label{x35}
  \begin{array}{c}
  \displaystyle{
 \mR_{12,\ti 1\ti 2}(z+N\frac{1}{M},\hbar)={\ti Q}_{\ti 1}^{-1}\,\mR_{12,\ti 1\ti
 2}(z,\hbar)\,
 {\ti Q}_{\ti 2}\,,
 }
 \\ \ \\
  \displaystyle{
 \mR_{12,\ti 1\ti 2}(z+N\frac{\tau}{M},\hbar)=\exp(-2\pi\imath N\frac{\hbar}{M})\,{\ti\Lambda}_{\ti 1}^{-1}\,\mR_{12,\ti 1\ti
 2}(z,\hbar)\,
 {\ti \Lambda}_{\ti 2}
 }
 \end{array}
 \eq
and
  \beq\label{x36}
  \begin{array}{c}
  \displaystyle{
 \mR_{12,\ti 1\ti 2}(z,\hbar+\frac{1}{N})=Q_1^{-1}{\ti Q}_{\ti 1}^{-1}\,\mR_{12,\ti 1\ti
 2}(z,\hbar)\,
  {\ti Q}_{\ti 1}\, Q_2\,,
 }
 \\ \ \\
  \displaystyle{
 \mR_{12,\ti 1\ti 2}(z,\hbar+\frac{\tau}{N})=\exp(-2\pi\imath \frac{z}{N})\,\Lambda_1^{-1}{\ti \Lambda}_{\ti 1}^{-1}\,\mR_{12,\ti 1\ti
 2}(z,\hbar)\,
{\ti \Lambda}_{\ti 1}\, \Lambda_2\,.
 }
 \end{array}
 \eq
The proof of (\ref{x35})-(\ref{x36}) is similar to the one given for
(\ref{x752}). For example, to get  (\ref{x36}) consider
 $$
 (T_\ga)_1^{-1}\, \mR_{12,\ti 1\ti 2}(z,\hbar) (T_\ga)_2=\sum\limits_{\al\in\, {\mathbb Z}_N\times {\mathbb Z}_N}
   \sum\limits_{\ti\al\in\, {\mathbb Z}_M\times {\mathbb
   Z}_M} \Phi_{\al,\ti\al}(z,\hbar)\,T_{\al-\ga}\otimes T_{\ga-\al}\otimes{\ti T}_{\ti\al}\otimes {\ti
   T}_{-\ti\al}\,.
 $$
Then, shift the summation index as $\al\rightarrow \al+\ga$ and
notice that
 $$\Phi_{\al+\ga,\ti\al}(z,\hbar)=\exp(2\pi\imath
z\p_\tau\om_\ga) {\ti\kappa}^2_{\ga,\ti\al}
\Phi_{\al,\ti\al}(z,\hbar+\om_\ga)\,.
 $$
Plugging $\ga=(1,0)$ (or $\ga=(0,1)$) into the obtained expression
one gets the upper (or the lower) line of (\ref{x36}).

\noindent {\bf Quadratic relations.} Let us prove that the symmetric
$R$-matrix obeys the relations similar to (\ref{z01}), (\ref{x04})
and (\ref{x05}).

 \begin{predl}
The symmetric $R$-matrix (\ref{x71}) satisfies the following set of
properties:

\noindent 1. Skew-symmetry:
  \beq\label{x741}
  \begin{array}{c}
  \displaystyle{
\mR_{21,\ti 2\ti 1}(-z,-\hbar)=-\mR_{12,\ti 1\ti 2}(z,\hbar)\,.
 }
 \end{array}
 \eq

\noindent 2. Associative Yang-Baxter equation:
  \beq\label{x742}
  \begin{array}{c}
  \displaystyle{
\mR_{12,\ti 1\ti 2}\, \mR_{23,\ti 3\ti 2}=\mR_{13,\ti 3\ti 2}\,
\mR_{12,\ti 1\ti 3} + \mR_{23,\ti 3\ti 1}\,\mR_{13,\ti 1\ti 2}
 }
 \end{array}
 \eq
 where $\mR_{ab,\ti a\ti b}=\mR_{ab,\ti a\ti b}(z_{a}-z_b,\hbar_{\ti a}-\hbar_{\ti b})$.

\noindent 3. Unitarity:
  \beq\label{x743}
  \begin{array}{c}
  \displaystyle{
\mR_{12,\ti 1\ti 2}\,\mR_{21,\ti 1\ti 2}=1_N\otimes 1_N\otimes {\ti
1}_M\otimes {\ti 1}_M\,N^2M^2(\wp(N\hbar)-\wp(Mz))\,.
 }
 \end{array}
 \eq
or
  \beq\label{x7431}
  \begin{array}{c}
  \displaystyle{
\mR_{12,\ti 1\ti 2}\,\mR_{12,\ti 2\ti 1}=1_N\otimes 1_N\otimes {\ti
1}_M\otimes {\ti 1}_M\,N^2M^2(\wp(Mz)-\wp(N\hbar))\,.
 }
 \end{array}
 \eq
 \end{predl}
\noindent\underline{\emph{Proof:}}
The first property simply follows from the definition
(\ref{x71})-(\ref{x72}):
 $$
\mR_{21,\ti 2\ti 1}(-z,-\hbar)=\sum\limits_{\al\in\, {\mathbb
Z}_N\times {\mathbb Z}_N}
   \sum\limits_{\ti\al\in\, {\mathbb Z}_M\times {\mathbb
   Z}_M} \Phi_{\al,\ti\al}(-z,-\hbar)\,T_{-\al}\otimes T_{\al}\otimes{\ti T}_{-\ti\al}\otimes {\ti
   T}_{\ti\al}
 $$
Changing summation indices as $\al\rightarrow-\al$,
$\ti\al\rightarrow-\ti\al$ and using
$\Phi_{-\al,-\ti\al}(-z,-\hbar)=-\Phi_{\al,\ti\al}(z,\hbar)$ we get
(\ref{x741}).

The second property follows from (\ref{a171}) (for both $\kappa$ and
$\ti\kappa$) and the Fay identity (\ref{aa904}) written in terms
functions (\ref{x72}) as
  \beq\label{x744}
  \begin{array}{l}
  \displaystyle{
\Phi_{\al,\ti\al}(z,\hbar)\,\Phi_{\be,\ti\be}(w,\eta)= }
 \\ \ \\
  \displaystyle{
 =\Phi_{\be,\ti\al+\ti\be}(z+w,\eta)\,\Phi_{\al-\be,\ti\al}(z,\hbar-\eta)
  +\Phi_{\be-\al,\ti\be}(w,\eta-\hbar)\,
  \Phi_{\al,\ti\al+\ti\be}(z+w,\hbar)
 }
 \end{array}
 \eq
with $z=z_1-z_2$, $w=z_2-z_3$, $\hbar=\hbar_{\ti 1}-\hbar_{\ti 2}$
and $\eta=\hbar_{\ti 3}-\hbar_{\ti 2}$.

To prove the unitarity (\ref{x743}) let us write its l.h.s.
explicitly
 $$
\mR_{12,\ti 1\ti 2}(z,\hbar)\,\mR_{21,\ti 1\ti
2}(-z,\hbar)=\sum\limits_{\al,\be,\ti\al,\ti\be}\Phi_{\al,\ti\al}(z,\hbar)\,\Phi_{\be,\ti\be}(-z,\hbar)\,
 \kappa_{\al,-\be}^2 {\ti\kappa}^2_{\ti\al,\ti\be}\,
 T_{\al-\be}\otimes T_{\be-\al}\otimes {\ti
 T}_{\ti\al+\ti\be}\otimes{\ti T}_{-\ti\al-\ti\be}
 $$
and subdivide this sum into four parts as
 $$
\sum\limits_{\al,\be,\ti\al,\ti\be}=\sum\limits_{\al=\be\,,\ti\al=-\ti\be}+
\sum\limits_{\al\neq\be\,,\ti\al=-\ti\be}+\sum\limits_{\al=\be\,,\ti\al\neq-\ti\be}+
\sum\limits_{\al\neq\be\,,\ti\al\neq-\ti\be}
 $$
The first sum reproduces the answer due to (\ref{aa906}) and
(\ref{aa9081})\footnote{We also use here that $\sum\limits_{\ti\al}
\wp(z+N{\ti\om}_\al)$=$\sum\limits_{\ti\al} \wp(z+{\ti\om}_\al)$ for
coprime $N$ and $M$.}:
 $$
\sum\limits_{\al=\be\,,\ti\al=-\ti\be}
\Phi_{\al,\ti\al}(z,\hbar)\,\Phi_{\be,\ti\be}(-z,\hbar)=\sum\limits_{\al}\sum\limits_{\ti\al}
\wp(\hbar+\om_\al)-\wp(z+N{\ti\om}_\al)=N^2M^2(\wp(N\hbar)-\wp(Mz))\,.
 $$
Each of three other sums equals zero. The vanishing of the second
and the third sum is proved in the same way as it is made for
unitarity of $R^\hbar(z)$ in the Appendix. Consider the fourth sum.
Denote $\ga=\al-\be$, $\ti\ga=\ti\al+\ti\be$. Then, using
(\ref{x744})
 $$
\sum\limits_{\al\neq\be\,,\ti\al\neq-\ti\be}=\sum\limits_{\ga\neq
0,\ti\ga\neq 0} T_{\ga}\otimes T_{-\ga}\otimes {\ti
 T}_{\ti\ga}\otimes{\ti T}_{-\ti\ga}\times
 $$
 $$
\times \left( \sum\limits_{\be,\ti\al} \kappa_{\be,\ga}^2\,
{\ti\kappa}^2_{\ti\al,\ti\ga}\,\Phi_{\be,\ti\ga}(0,\hbar)\,\Phi_{\ga,\ti\al}(z,0)+
 \sum\limits_{\al,\ti\be} \kappa_{\al,\ga}^2\,
{\ti\kappa}^2_{\ti\ga,\ti\be}\,\Phi_{\al,\ti\ga}(0,\hbar)\,\Phi_{-\ga,\ti\be}(-z,0)
\right)\,.
 $$
The latter two terms cancel each other after changing notation
$\al:=\be$, $\ti\be:=-\ti\al$ in the last term. This finishes the
proof of (\ref{x743}). At last, (\ref{x7431}) follows from
(\ref{x743}) and (\ref{x741}). $\blacksquare$

\noindent {\bf Cubic relations.} The next step is to derive cubic
relations of (\ref{x21})-(\ref{x24}) type. Recall that (\ref{x23})
is the most general in that list. In a similar way one can obtain
its analogue for the symmetric $R$-matrix (\ref{x71}). First,
multiply (\ref{x742}) by $\mR_{23,\ti 1\ti 3}$ from the left:
 $$
\mR_{23,\ti 1\ti 3}\,\mR_{12,\ti 1\ti 2}\, \mR_{23,\ti 3\ti
2}=\mR_{23,\ti 1\ti 3}\,\mR_{13,\ti 3\ti 2}\, \mR_{12,\ti 1\ti 3} +
\mR_{23,\ti 1\ti 3}\,\mR_{23,\ti 3\ti 1}\,\mR_{13,\ti 1\ti 2}\,.
 $$
Second, change indices $2\leftrightarrow 3$ in (\ref{x742}) and
multiply it by $\mR_{23,\ti 3\ti 2}$ from the right:
 $$
\mR_{13,\ti 1\ti 2}\, \mR_{32,\ti 3\ti 2}\,\mR_{23,\ti 3\ti
2}=\mR_{12,\ti 3\ti 2}\, \mR_{13,\ti 1\ti 3}\,\mR_{23,\ti 3\ti 2} +
\mR_{32,\ti 3\ti 1}\,\mR_{12,\ti 1\ti 2}\,\mR_{23,\ti 3\ti 2}\,.
 $$
Combining these two equations and using the properties of
skew-symmetry (\ref{x741}) and unitarity (\ref{x743})-(\ref{x7431})
we get
  \beq\label{x745}
  \begin{array}{l}
  \displaystyle{
\mR_{12,\ti 3\ti 2}\, \mR_{13,\ti 1\ti 3}\,\mR_{23,\ti 3\ti
2}=\mR_{23,\ti 1\ti 3}\,\mR_{13,\ti 3\ti 2}\, \mR_{12,\ti 1\ti
3}+N^2M^2\,\mR_{13,\ti 1\ti 2}\Big( \wp(N\hbar_{\ti 3\ti 2}) -
\wp(N\hbar_{\ti 1\ti 3}) \Big)\,,
 }
 \end{array}
 \eq
where $\hbar_{\ti a\ti b}=\hbar_{\ti a}-\hbar_{\ti b}$. When $M=1$
(\ref{x745}) coincides with (\ref{x23}), where $\hbar=\hbar_{\ti
1}-\hbar_{\ti 3}$ and $\eta=\hbar_{\ti 3}-\hbar_{\ti 2}$.

In particular case of $\hbar_{\ti 3}=(\hbar_{\ti 1}+\hbar_{\ti
2})/2$ we get Yang-Baxter like equation
  \beq\label{x746}
  \begin{array}{l}
  \displaystyle{
\mR_{12,\ti 3\ti 2}(z_{12},\hbar)\, \mR_{13,\ti 1\ti
3}(z_{13},\hbar)\,\mR_{23,\ti 3\ti 2}(z_{23},\hbar)=\mR_{23,\ti 1\ti
3}(z_{23},\hbar)\,\mR_{13,\ti 3\ti 2}(z_{13},\hbar)\, \mR_{12,\ti
1\ti 3}(z_{12},\hbar)
 }
 \end{array}
 \eq
with $\hbar={\ti\hbar}_{\ti 1\ti 3}={\ti\hbar}_{\ti 3\ti 2}$.

Another particular case ${\ti\hbar}_{\ti 1}={\ti\hbar}_{\ti 2}$ in
(\ref{x745}) leads to analogue of (\ref{x22}):
  \beq\label{x747}
  \begin{array}{c}
  \displaystyle{
\mR_{21,\ti 2\ti 3}(z_{21},\hbar)\, \mR_{13,\ti 1\ti
3}(z_{13},\hbar)\,\mR_{32,\ti 2\ti 3}(z_{32},\hbar)+\mR_{23,\ti 1\ti
3}(z_{23},\hbar)\,\mR_{31,\ti 2\ti 3}(z_{31},\hbar)\, \mR_{12,\ti
1\ti 3}(z_{12},\hbar)=
 }
 \\ \ \\
  \displaystyle{
=-1_N\otimes 1_N\otimes {\ti P}_{\ti 1\ti 3}\,\,
N^3M^2\wp'(N\hbar)\,.
 }
 \end{array}
 \eq
A similar derivation leads to the cubic equation of (\ref{x745})
type with $\hbar$ and $z$ variables interchanged:
  \beq\label{x7471}
  \begin{array}{l}
  \displaystyle{
\mR_{32,\ti 1\ti 2}\, \mR_{13,\ti 1\ti 3}\,\mR_{32,\ti 2\ti
3}=\mR_{13,\ti 2\ti 3}\,\mR_{32,\ti 1\ti 3}\, \mR_{13,\ti 1\ti
2}+N^2M^2\,\mR_{12,\ti 1\ti 3}\Big( \wp(Mz_{23}) - \wp(Mz_{13})
\Big)\,,
 }
 \end{array}
 \eq
where $z_{ab}=z_a-z_b$.

Now we can easily obtain the cubic Yang-Baxter like equations of
(\ref{x24}) type. It is based on the skew-symmetry of
$\wp(N\hbar_{\ti 3\ti 2}) - \wp(N\hbar_{\ti 1\ti 3})$ with respect
to interchanging of the arguments only. Therefore, from (\ref{x745})
we have:
  $$
  \begin{array}{l}
  \displaystyle{
\mR_{12,\ti 3\ti 2}(z_{12},\hbar_{\ti 3\ti 2})\, \mR_{13,\ti 1\ti
3}(z_{13},\hbar_{\ti 1\ti 3})\,\mR_{23,\ti 3\ti 2}(z_{23},\hbar_{\ti
3\ti 2})+\mR_{12,\ti 3\ti 2}(z_{12},\hbar_{\ti 1\ti 3})\,
\mR_{13,\ti 1\ti 3}(z_{13},\hbar_{\ti 3\ti 2})\,\mR_{23,\ti 3\ti
2}(z_{23},\hbar_{\ti 1\ti 3})
 }
 \\ \ \\
  \displaystyle{
=\mR_{23,\ti 1\ti 3}(z_{23},\hbar_{\ti 1\ti 3})\mR_{13,\ti 3\ti
2}(z_{13},\hbar_{\ti 3\ti 2}) \mR_{12,\ti 1\ti 3}(z_{12},\hbar_{\ti
1\ti 3})+\mR_{23,\ti 1\ti 3}(z_{23},\hbar_{\ti 3\ti 2})\mR_{13,\ti
3\ti 2}(z_{13},\hbar_{\ti 1\ti 3}) \mR_{12,\ti 1\ti
3}(z_{12},\hbar_{\ti 3\ti 2})
 }
 \end{array}
 $$
In the same manner from (\ref{x7471}) we get:
 $$
  \begin{array}{l}
  \displaystyle{
\mR_{32,\ti 1\ti 2}(z_{32},\hbar_{\ti 1\ti 2})\, \mR_{13,\ti 1\ti
3}(z_{13},\hbar_{\ti 1\ti 3})\,\mR_{32,\ti 2\ti 3}(z_{32},\hbar_{\ti
2\ti 3})+\mR_{32,\ti 1\ti 2}(z_{13},\hbar_{\ti 1\ti 2})\,
\mR_{32,\ti 1\ti 3}(z_{13},\hbar_{\ti 1\ti 3})\,\mR_{32,\ti 2\ti
3}(z_{13},\hbar_{\ti 2\ti 3})
 }
 \\ \ \\
  \displaystyle{
=\mR_{13,\ti 2\ti 3}(z_{13},\hbar_{\ti 2\ti 3})\mR_{32,\ti 1\ti
3}(z_{32},\hbar_{\ti 1\ti 3}) \mR_{13,\ti 1\ti 2}(z_{13},\hbar_{\ti
1\ti 2})+\mR_{13,\ti 2\ti 3}(z_{32},\hbar_{\ti 2\ti 3})\mR_{32,\ti
1\ti 3}(z_{13},\hbar_{\ti 1\ti 3}) \mR_{13,\ti 1\ti
2}(z_{32},\hbar_{\ti 1\ti 2})
 }
 \end{array}
 $$

\noindent {\bf Relation to Baxter-Belavin ${\rm GL}_{NM}$
$R$-matrix.} Multiply the symmetric $R$-matrix by $N P_{12}\otimes
{\ti 1}_M\otimes {\ti 1}_M$:
 $$
\mR_{12,\ti 1\ti 2}(z,\hbar)\, N P_{12}\otimes {\ti 1}_M\otimes {\ti
1}_M=\sum\limits_{\al,\be\in\, {\mathbb Z}_N\times {\mathbb Z}_N}
   \sum\limits_{\ti\al\in\, {\mathbb Z}_M\times {\mathbb
   Z}_M} \Phi_{\al,\ti\al}(z,\hbar)\,T_\al T_\be\otimes T_{-\al}T_{-\be}\otimes{\ti T}_{\ti\al}\otimes {\ti
   T}_{-\ti\al}
 $$
 $$
=\sum\limits_{\al,\ga,\ti\al}\kappa_{\al,\ga}^2
\exp(N\hbar\,\p_\tau{\ti\om}_{\ti\al})
\vf_\al(z+N{\ti\om}_{\ti\al},\hbar+\om_\al)\, T_\ga\otimes
T_{-\ga}\otimes{\ti T}_{\ti\al}\otimes {\ti
   T}_{-\ti\al}\stackrel{(\ref{a910})}{=}
 $$
 $$
=N \sum\limits_{\ga,\ti\al}
\vf_{\ga+\ti\al}(N\hbar,\om_\ga+{\ti\om}_{\ti\al}+\frac{z}{N})\,
T_\ga\otimes T_{-\ga}\otimes{\ti T}_{\ti\al}\otimes {\ti
   T}_{-\ti\al}\,.
 $$
Since $M$ and $N$ are coprime the summation over
$\om_\ga+{\ti\om}_{\ti\al}$ is equivalent to the summation over
$\om_a$, where $a\in{\mathbb Z}_{MN}\times {\mathbb Z}_{MN}$.
Therefore, we get the expression which is the ${\rm GL}_{NM}$
Baxter-Belavin's $R$-matrix written in special basis.

\noindent {\bf Particular case $M=N$.} In this particular case the
function $\Phi_{\al,\ti\al}(z,\hbar)$ (\ref{x72}) is simplified.
Using (\ref{aa9031}) it is easy to get that
$$
\Phi_{\al,\ti\al}(z,\hbar)=\exp(2\pi\imath\frac{
{\ti\al}_1\al_2-{\ti\al}_2\al_1
}{N})\vf_\al(z,\om_\al+\hbar)=\kappa_{\al,\ti\al}^2\,\vf_\al(z,\om_\al+\hbar)\,.
$$





\section{Appendix}
\def\theequation{A.\arabic{equation}}
\setcounter{equation}{0}

\subsection{Elliptic functions}
 We deal with the following
elliptic functions:
 \beq\label{aa901}
 \begin{array}{c}
  \displaystyle{
 \phi(z,u)=\frac{\vth'(0)\vth(z+u)}{\vth(z)\vth(u)}\,,
  }
 \end{array}
 \eq
  \beq\label{aa902}
 \begin{array}{c}
  \displaystyle{
 E_1(z)=\frac{\vth'(z)}{\vth(z)}\,,\ \ \ \ \ E_2(z)=-\p_z E_1(z)=\wp(z)-\frac{1}{3}\frac{\vth'''(0)}{\vth'(0)}\,,
  }
 \end{array}
 \eq
where $\vth(z)$ is the Riemann theta-function
 \beq\label{aa903}
 \begin{array}{c}
  \displaystyle{
\vth(z)=\vth(z|\tau)=\displaystyle{\sum _{k\in \mathbb Z}} \exp
\left ( \pi \imath \tau (k+\frac{1}{2})^2 +2\pi \imath
(z+\frac{1}{2})(k+\frac{1}{2})\right )
  }
 \end{array}
 \eq
 which has simple zero at $z=0$, and $\wp(z)$ in (\ref{aa902}) is the Weierstrass $\wp$-function.

The function $\phi(x,y)=\phi(y,x)$ has the quasiperiodic properties
 \beq\label{aa9031}
 \begin{array}{c}
  \displaystyle{
\phi(x+1,y)=\phi(x,y)\,,\qquad \phi(x+\tau,y)=\exp(-2\pi\imath
y)\phi(x,y)
  }
 \end{array}
 \eq
and satisfies the Fay type identities:
  \beq\label{aa904}
  \begin{array}{c}
  \displaystyle{
 \phi(x,u)\phi(y,w)=\phi(x-y,u)\phi(y,u+w)+\phi(y-x,w)\phi(x,u+w)\,,
 }
 \end{array}
 \eq
 \beq\label{aa905}
  \begin{array}{c}
  \displaystyle{
 \phi(x,z)\phi(x,w)=\phi(x,z+w)(E_1(x)+E_1(z)+E_1(w) -E_1(x+z+w)
 )\,,
 }
 \end{array}
 \eq
 \beq\label{aa906}
  \begin{array}{c}
  \displaystyle{
 \phi(z,u)\phi(z,-u)=\wp(z)-\wp(u)=E_2(z)-E_2(u)\,.
 }
 \end{array}
 \eq
These relations lead to
 \beq\label{aa907}
  \begin{array}{c}
  \displaystyle{
 \phi(z,u_1-v)\phi(w,u_2+v)\phi(z-w,v)-\phi(z,u_2+v)\phi(w,u_1-v)\phi(z-w,u_1-u_2-v)
 }
 \\ \ \\
  \displaystyle{
 =\phi(z,u_1)\phi(w,u_2) ( E_1(v)-E_1(u_1-u_2-v)+E_1(u_1-v)-E_1(u_2+v) )
 }
 \end{array}
 \eq
and
 \beq\label{aa908}
  \begin{array}{c}
  \displaystyle{
 \phi(z,u-v)\phi(w,v)\phi(z-w,v)-\phi(z,v)\phi(w,u-v)\phi(z-w,u-v)
 }
 \\ \ \\
  \displaystyle{
 =\phi(z,u) ( \wp(v)-\wp(u-v) )\,.
 }
 \end{array}
 \eq
We also need the following set of functions:
  \beq\label{a912}
 \begin{array}{c}
  \displaystyle{
\vf_\al(z,\om_\al+\hbar)=\exp\left(2\pi\imath\frac{\al_2}{N}z\right)\phi(z,\om_\al+\hbar)\,,\qquad
\al\in{\mathbb Z}^{\times 2}\,.
  }
 \end{array}
 \eq
 These functions satisfy the set of identities
 (\ref{aa904})-(\ref{aa908}) because the exponential factors are canceled.

Averaging  $\wp$-function:
 \beq\label{aa9081}
  \begin{array}{c}
  \displaystyle{
\sum\limits_{\al\in{\mathbb Z}_N\times {\mathbb Z}_N}
\wp(\om_\al+\hbar)=N^2\wp(N\hbar)
 }
 \end{array}
 \eq

The following formulae (of finite Fourier transformation type) are
also useful in derivation of $R$-matrix identities (see
\cite{LOZ10}):
 \beq\label{aa909}
 \begin{array}{c}
  \displaystyle{
 \frac{1}{N}\sum\limits_{\al\in{\mathbb Z}_N\times {\mathbb Z}_N} \kappa_{\al,\ga}^2\,
\vf_\al(N\hbar,\om_\al+\frac{z}{N})=\vf_\ga(z,\om_\ga+\hbar)\,,\ \
 \forall\,\ga\in{\mathbb Z}^{\times 2}
  }
 \end{array}
 \eq
 or (exchanging the arguments)
  \beq\label{a910}
 \begin{array}{c}
  \displaystyle{
 \frac{1}{N}\sum\limits_{\al\in{\mathbb Z}_N\times {\mathbb Z}_N} \kappa_{\al,\ga}^2\,
\vf_\al(z,\om_\al+\hbar)=\vf_\ga(N\hbar,\om_\ga+\frac{z}{N})\,,\ \
 \forall\,\ga\in{\mathbb Z}^{\times 2}\,,
  }
 \end{array}
 \eq
where $\kappa_{\al,\be}$ is from (\ref{a17}).

\subsection{Baxter-Belavin $R$-matrix}
\noindent {\bf Finite-dimensional representation of the Heisenberg
group.} Consider the following $N\times N$ matrices:
  \beq\label{a14}
 \begin{array}{c}
  \displaystyle{
Q,\Lambda\in \hbox{Mat}(N,\mathbb C):\ \ \
Q_{kl}=\delta_{kl}\exp(\frac{2\pi
 \imath}{N}k)\,,\ \ \ \Lambda_{kl}=\delta_{k-l+1=0\,{\hbox{\tiny{mod}}}
 N}\,,\ \ k,l=1,...,N\,,
 }
 \end{array}
 \eq
  \beq\label{a141}
 \begin{array}{c}
  \displaystyle{
Q^N=\Lambda^N=1\,,
 }
 \end{array}
 \eq
  \beq\label{a15}
 \begin{array}{c}
  \displaystyle{
\exp(2\pi
\imath\frac{\ga_1\ga_2}{N})\,Q^{\ga_1}\Lambda^{\ga_2}=\Lambda^{\ga_2}
Q^{\ga_1}\,,\ \ \ga_1\,,\ga_2\in\mathbb Z\,.
 }
 \end{array}
 \eq
Introduce
  \beq\label{a16}
 \begin{array}{c}
  \displaystyle{
T_\ga:=T_{\ga_1\ga_2}=\exp(\pi
\imath\frac{\ga_1\ga_2}{N})\,Q^{\ga_1}\Lambda^{\ga_2}\,,\ \
\ga=(\ga_1,\ga_2)\in{\mathbb Z}^{\times 2}\,.
 }
 \end{array}
 \eq
The subset of $T_\ga$ with $\ga_1\,,\ga_2=0,...,N-1$ is the basis in
$\hbox{Mat}(N,\mathbb C)$. From (\ref{a15}) we have
  \beq\label{a17}
 \begin{array}{c}
  \displaystyle{
T_\al T_\be=\kappa_{\al,\be} T_{\al+\be}\,,\ \ \
\kappa_{\al,\be}=\exp\left(\frac{\pi \imath}{N}(\be_1
\al_2-\be_2\al_1)\right)\,,
 }
 \end{array}
 \eq
where $\al+\be=(\al_1+\be_1,\al_2+\be_2)$. Obviously,
  \beq\label{a171}
 \begin{array}{c}
  \displaystyle{
\kappa_{\al,\be}\kappa_{\be,\al}=1\,,\quad
\kappa_{\al,-\be}=\kappa_{-\al,\be}=\kappa_{\be,\al}\,,\quad
\kappa_{\al,\be}=\kappa_{\al+\be,\be}=\kappa_{\al,\be+\al}\,.
 }
 \end{array}
 \eq
Notice also that
 \beq\label{a503}
 \begin{array}{c}
  \displaystyle{
\sum\limits_{\al} \kappa_{\al,\ga}^2=N^2\delta_{\ga,\,0}\,,
  }
 \end{array}
 \eq
where $0=(0,0)$.

For $N=2$ the basis (\ref{a16}) in $2\times 2$ matrices coincides
(up to signs) with the Pauli matrices basis $\sigma_j$ endowed with
$\sigma_0=1_{2\times 2}$. In this case (\ref{a17}) can be re-written
as $\sigma_j\sigma_k=\sigma_0
\delta_{jk}+\imath\varepsilon_{jkn}\sigma_n$ together with
$\sigma_j\sigma_0=\sigma_0\sigma_j=\sigma_j$ and
$\sigma_0^2=\sigma_0$.

\noindent {\bf $R$-matrix} is defined as\footnote{Let us mention
that original definition \cite{Belavin} differs from (\ref{a18}). It
was defined in terms of theta functions with characteristics $1/N$
and normalized as $R^\hbar_{12}R^\hbar_{21}=1\otimes 1$. Its
representation in the standard basis of $\Mat$ $\{
\left(E_{ij}\right)_{kl}=\delta_{ik}\delta_{jl} \}$ was suggested in
\cite{RicheyT}. }
 \beq\label{a18}
 \begin{array}{c}
  \displaystyle{
R^\hbar_{12}(u)=\sum\limits_{\al\in\, {\mathbb Z}_N\times {\mathbb
Z}_N}\vf_\al(u,\om_\al+\hbar)\,T_\al\otimes T_{-\al}\in
\hbox{Mat}(N,\mathbb C)^{\otimes 2}\,,
 }
 \end{array}
 \eq
 where
 \beq\label{a19}
 \begin{array}{c}
  \displaystyle{
\vf_\al(u,\om_\al+\hbar)=\exp(2\pi\imath
\frac{\al_2}{N}u)\,\phi(u,\om_\al+\hbar)\,,\ \ \
\om_\al=\frac{\al_1+\al_2\tau}{N}\,.
 }
 \end{array}
 \eq
Writing $R^\hbar_{ab}(z)$ we mean that $T_\al\otimes T_{-\al}$ in
 (\ref{a18}) is replaces by  $1\otimes ...  1\otimes T_{\al}\otimes 1... 1\otimes T_{-\al}\otimes 1...
 \otimes1$ in such a way that $T_\al$ and $T_{-\al}$ acts in the $a$-th and $b$-th components of $\Mat^{\otimes n\geq 3}$.
  If the number of components equals $n$ then
 $R_{ab}^\hbar\in {\rm Mat}(N^{n},\mathbb C)$.
With the above definition we get the correct relation
 \beq\label{a190}
 \begin{array}{c}
  \displaystyle{
R_{ba}^\hbar(z)=P_{ab}\, R_{ab}^\hbar(-z)\, P_{ab}\,,
 }
 \end{array}
 \eq
where $P_{12}$ is the permutation operator in $\Mat^{\otimes 2}$.
Notice also that $P_{12}$ has the following form in the basis
(\ref{a16}):
 \beq\label{a191}
 \begin{array}{c}
  \displaystyle{
\sum\limits_\al T_\al\otimes T_{-\al}=NP_{12}\,,
 }
 \end{array}
 \eq

The classical limit $\hbar\rightarrow 0$ is defined as
  \beq\label{x051}
  \begin{array}{c}
  \displaystyle{
 R^\hbar_{12}(z)=\hbar^{-1}1\otimes
 1+r_{12}(z)+\hbar\,m_{12}(z)+O(\hbar^2)\,.
 }
 \end{array}
 \eq
 Then the quantum Yang-Baxter equation (\ref{yb}) provides the classical
 one
 \beq\label{x052}
 \begin{array}{c}
  \displaystyle{
[r_{ab},r_{ac}]+[r_{ac},r_{bc}]+[r_{ab},r_{bc}]=0\,,\ \
r_{ab}=r_{ab}(z_a-z_b)\,.
 }
 \end{array}
 \eq
Condition (\ref{x04}) leads to
 \beq\label{x053}
 \begin{array}{c}
  \displaystyle{
r_{ab}=-r_{ba}\,, \ \ m_{ab}=m_{ba}\,,
 }
 \end{array}
 \eq
while (\ref{x05}) gives expresses $m_{ab}$ in terms of $r_{ab}$:
 \beq\label{x054}
 \begin{array}{c}
  \displaystyle{
2m_{12}(z)=r_{12}^2(z)-1\otimes 1 N^2\wp(z)\,.
 }
 \end{array}
 \eq
The answer is similar to $\hbar^1$ term in the local expansion of
the Kronecker function near $\hbar=0$:
 $$
\phi(\hbar,z)=\hbar^{-1}+E_1(z)+\hbar\,(E_1^2(z)-\wp(z))/2+O(\hbar^2)\,.
 $$
At the same time the Fay identity for $R$-matrix (\ref{x03}) gives
 \beq\label{x055}
 \begin{array}{c}
  \displaystyle{
r_{ab}\,r_{ac}-r_{bc}\,r_{ab}+r_{ac}\,r_{bc}=m_{ab}+m_{bc}+m_{ac}\,.
 }
 \end{array}
 \eq
Plugging $m_{ab}$ from (\ref{x054}) into  (\ref{x055}) one gets the
non-abelian analogue of
 $$
(E_1(z_a-z_b)+E_1(z_b-z_c)+E_1(z_c-z_a))^2=\wp(z_a-z_b)+\wp(z_b-z_c)+\wp(z_c-z_a).
 $$

In the end let us prove the unitarity property (\ref{x05}) (this
proof was skipped in \cite{LOZ9,LOZ10}).

\noindent {\bf Unitarity.} \underline{Proof of (\ref{x05}).} It is
convenient to rewrite (\ref{x05}) using (\ref{x04}) as
$$R_{12}^\hbar(z) R_{12}^{-\hbar}(z)=-N^2(\wp(N\hbar)-\wp(z))\,
1\otimes 1\,.$$
 The proof is achieved by direct calculation:
 $$
R_{12}^\hbar(z) R_{12}^{-\hbar}(z)=\sum\limits_{\al,\be} T_\al
T_\be\otimes T_{-\al}T_{-\be}
\vf_\al(z,\om_\al+\hbar)\vf_\be(z,\om_\be-\hbar)=\sum\limits_{\al+\be=0}+\sum\limits_{\al+\be\neq0}
 $$
The first term reproduces the answer due to (\ref{aa906}) and
$\sum_\al \wp(\om_\al+\hbar)=N^2\wp(N\hbar)$. So we need to prove
that the second term (where $\al+\be\neq 0$) equals 0:
 $$
\sum\limits_{\al+\be\neq0}\stackrel{(\ref{a17}),(\ref{aa905})}{=}\sum\limits_{\ga\neq
0} \sum\limits_{\al+\be=\ga} T_\ga\otimes
T_{-\ga}\,\kappa_{\al,\be}^2
\vf_\ga(z,\om_\ga)(E_1(z)+E_1(\om_\al+\hbar)+E_1(\om_\be-\hbar)-E_1(z+\om_\ga))
 $$
The first and the last terms in the brackets are independent of
$\al,\be$. Each of them gives 0 since
$\kappa_{\al,\be}=\kappa_{\al,\al+\be}=\kappa_{\al,\ga}$ and
$\sum_\al \kappa_{\al,\ga}^2=N^2\delta_{\ga,0}$ but the sum does not
contain $\ga=0$ term. Finally, we are left with
 $$
\sum\limits_{\al+\be=\ga} \,\kappa_{\al,\be}^2
(E_1(\om_\al+\hbar)+E_1(\om_\be-\hbar))
=\sum\limits_{\al}\kappa_{\al,\ga}^2(E_1(\om_\al+\hbar)+E_1(\om_\ga\!-\!\om_\al-\hbar))=0
 $$
because the summation index of the second term can be shifted as
$\al\rightarrow \al+\gamma$, and due to $E_1(x)=-E_1(-x)$.

More details and properties can be found in \cite{RicheyT,Hasegawa}
and \cite{LOZ10}.

\subsection{${\rm GL}_N$ Sklyanin algebra}
The Sklyanin algebra \cite{Sklyanin} has extension to ${\rm GL}_N$.
One possible approach is given in \cite{OF}. The most natural way is
to use the exchange relations (as in \cite{Sklyanin}).
 \begin{predl}
Consider the Lax operator (\ref{x52}) and $R$-matrix (\ref{x06}).
The quantum exchange relations
$$R_{12}^\hbar(z-w)\hL_1^\hbar(z)\hL_2^\hbar(w)=\hL_2^\hbar(w)\hL_1^\hbar(z)R_{12}^\hbar(z-w)$$
are equivalent to $N^2\times N^2$ relations numbered by
$\al,\be\in{\mathbb Z}_N^{\times 2}$ (for the component
$T_\al\otimes T_\be$):
 \beq\label{x61}
 \begin{array}{c}
  \displaystyle{
\sum\limits_\ga \kappa_{\ga,\al-\be}
\hS_{\al-\ga} \hS_{\be+\ga}\,{\rm f}^\hbar_{\al,\be,\ga}=0\,,
 }
 \end{array}
 \eq
 where the structure constants ${\rm
f}^\hbar_{\al,\be,\ga}$ are given by
 \beq\label{x62}
 \begin{array}{l}
  \displaystyle{
{\rm for}\ \be\neq 0:\quad {\rm
f}^\hbar_{\al,\be,\ga}=E_1(\om_\ga+\hbar)-E_1(\om_{\al-\be-\ga}+\hbar)+E_1(\om_{\al-\ga}+\hbar)-E_1(\om_{\be+\ga}+\hbar)\,,
 }
 \\ \ \\
  \displaystyle{
{\rm for}\ \be=0:\quad {\rm
f}^\hbar_{\al,0,\ga}=\wp(\om_\ga+\hbar)-\wp(\om_{\al-\ga}+\hbar)\,.
 }
 \end{array}
 \eq
 \end{predl}
For $N=2$ relations (\ref{x61}) reproduce ${\rm GL}_2$ Sklyanin
algebra in its original form \cite{Sklyanin}. Indeed, for
$\al,\be\neq 0$ and $\al\neq \be$ the structure constants are
related: ${\rm f}^\hbar_{\al,\be,\al-\be}={\rm
f}^\hbar_{\al,\be,-\be}=-{\rm f}^\hbar_{\al,\be,0}$ and ${\rm
f}^\hbar_{\al,\be,\al}={\rm
f}^\hbar_{\al,\be,0}$. 
Therefore,
 \beq\label{x63}
 \begin{array}{c}
  \displaystyle{
[\hS_\al,\hS_\be]_-=\kappa_{\al,\be}[\hS_{\al+\be},\hS_0]_+\,.
 }
 \end{array}
 \eq
Here we also used that for $\al,\be\neq 0$ and $\al\neq \be$
$\kappa_{\al,\be}=\pm\imath$, i.e.
$\kappa_{\al,\be}=-\kappa_{\be,\al}$. To get the second commutation
relation introduce
 \beq\label{x64}
 \begin{array}{c}
  \displaystyle{
\wp^\hbar_\al=\wp(\hbar+\om_\al)-\wp(\hbar)
 }
 \end{array}
 \eq
and notice that ${\rm f}^\hbar_{\al,0,\al}=-{\rm
f}^\hbar_{\al,0,0}$, ${\rm f}^\hbar_{\al,0,\be}=-{\rm
f}^\hbar_{\al,0,\al-\be}$. Then
 \beq\label{x65}
 \begin{array}{c}
  \displaystyle{
\wp^\hbar_\al\,[\hS_0,\hS_{\al+\be}]_-=-\kappa_{\al,\be}(\wp^\hbar_{\al}-\wp^\hbar_{\be})[\hS_{\al},\hS_\be]_+\,.
 }
 \end{array}
 \eq
 for some $\be\neq \al,0$. Notice also that the same commutation relation appears from $\al=0$, $\be\neq
 0$ component of (\ref{x61}) but in a different form:
$$
K^\hbar_{\al+\be}\,[\hS_0,\hS_{\al+\be}]_-=\kappa_{\al,\be}(K^\hbar_{\al}-K^\hbar_{\be})[\hS_{\al},\hS_\be]_+\,,
$$
 where $K^\hbar_\al=E_1(\hbar+\om_\al)-E_1(\hbar)-E_1(\om_\al)$. The
 latter is the same as (\ref{x65}) due to
 $$\frac{K^\hbar_{\al}-K^\hbar_{\be}}{K^\hbar_{\al+\be}}=-\frac{\wp^\hbar_\al-\wp^\hbar_\be}{\wp^\hbar_{\al+\be}}$$
in ${\rm GL}_2$ case.

     \renewcommand{\refname}{{\normalsize{References}}}


 \begin{small}

 \end{small}

\end{document}